\documentclass[12pt]{article}
\usepackage{epsf}
\usepackage{amsmath,amssymb}

\usepackage[dvips]{graphicx}

\usepackage{comment}
\usepackage{bm}

\usepackage{graphicx}
\usepackage{listings}
\usepackage{xcolor}
\usepackage{bm}
\usepackage{authblk}
\begin{document}

\newcommand{\lsim}{\stackrel{<}{_\sim}}
\newcommand{\gsim}{\stackrel{>}{_\sim}}
\newcommand{\mathhyphen}{\mathchar"712D}

\newcommand{\rem}[1]{{$\spadesuit$\bf #1$\spadesuit$}}

\renewcommand{\theequation}{\thesection.\arabic{equation}}

\renewcommand{\thefootnote}{\fnsymbol{footnote}}
\setcounter{footnote}{0}

\begin{titlepage}

\def\thefootnote{\fnsymbol{footnote}}

\begin{center}

\hfill June, 2024\\

\vskip .5in

{\Large \bf

  On the Metastability of Quantum Fields
  \\
  in Thermal Bath
  \\
  
}

\vskip .5in

{\large
  Zhiyi Fan and Takeo Moroi
}

\vskip .5in

{\em
Department of Physics, The University of Tokyo, Tokyo 113-0033, Japan
}

\end{center}
\vskip .5in

\begin{abstract}

We investigate the metastability of scalar fields in quantum field theories at finite temperature, focusing on a detailed understanding of the bounce solution. At finite temperature, the bounce solution depends on two variables: the Euclidean time $\tau$ and the spatial radial distance $r$, and it is periodic in the $\tau$ direction. We propose a novel method to determine the bounce that describes transitions in a thermal bath, suitable for numerical calculations. Two types of bounces exist for transitions in a thermal bath: $\tau$-dependent and $\tau$-independent bounces. We apply our method to compute these bounces in several models, including both thin-wall and thick-wall scenarios, to examine their properties. Specifically, we evaluate the critical temperature below which the $\tau$-independent bounce becomes destabilized due to fluctuations, rendering it irrelevant. We demonstrate that in the thick-wall case, the $\tau$-dependent bounce smoothly transitions into the $\tau$-independent one as temperature increases, whereas in the thin-wall case, the transition between the two types of bounces is discontinuous.

\end{abstract}

\end{titlepage}

\renewcommand{\thepage}{\arabic{page}}
\setcounter{page}{1}
\renewcommand{\thefootnote}{\#\arabic{footnote}}
\setcounter{footnote}{0}
\renewcommand{\theequation}{\thesection.\arabic{equation}}

\section{Introduction}
\label{sec:intro}
\setcounter{figure}{0}

Fates of metastable states in quantum field theories (QFTs) are of great interest. In a large class of particle-physics models, the stability of the ``vacuum'' (which is given by the minimum of the potential) is not guaranteed. This holds true even within the standard model of particle physics; the electroweak vacuum becomes unstable because the Higgs quartic coupling constant runs to negative at a high scale due to the renormalization group effect. Extensive studies have been conducted on the lifetime of the electroweak vacuum, revealing it to be considerably longer than the present age of the universe \cite{Degrassi:2012ry, Elias-Miro:2011sqh, Andreassen:2017rzq, Chigusa:2017dux, Chigusa:2018uuj}. Furthermore, many particle-physics models predict the existence of both false and true vacua; considerations on the metastable states often give us insights into physics beyond the standard model. Additionally, a transition from a metastable state to a stable one (i.e., a phase transition) may have occurred in the early universe, potentially leading to observable consequences such as gravitational waves. 

In the study of the transition of metastable states in QFT, a field configuration called ``bounce'' plays a pivotal role. The path integral for the transition process is dominated by the bounce configuration which is a solution of the Euclidean equation of motion (EoM). Transition from false vacuum to true vacuum at zero temperature was detailed in Refs.\ \cite{Coleman:1977py, Callan:1977pt, Coleman:1985rnk}. In $D$-dimensional space-time, the bounce is $O(D)$ symmetric at zero temperature \cite{Coleman:1977th, Blum:2016ipp} and the bounce amplitude depends solely on one variable, the radius from the center of the bounce. In such a case, a number of methods exist to calculate the bounce \cite{Claudson:1983et, Kusenko:1995jv, Kusenko:1996jn, Dasgupta:1996qu, Moreno:1998bq, John:1998ip, Cline:1998rc, Cline:1999wi, Konstandin:2006nd, Wainwright:2011kj, Akula:2016gpl, Masoumi:2016wot, Espinosa:2018hue, Jinno:2018dek, Espinosa:2018szu, Athron:2019nbd, Chigusa:2019wxb, Sato:2019axv, Hong:2023dan}; given a scalar potential which has a false vacuum, we can numerically calculate the bounce configuration with such methods to study its properties in detail. 

The primary focus of this paper is to examine the metastability of quantum fields at finite temperature; such a subject is of particular interest in the light of the applications to the phase transitions in the early universe. We follow the analysis conducted by Affleck \cite{Affleck:1980ac} (see also Refs.\ \cite{Linde:1980tt, Linde:1981zj, Garriga:1994ut}), which demonstrated that the transition rate at finite temperature is proportional to ${\rm Im}F$, where $F$ represents the free energy:
\begin{align}
  F \equiv -\frac{1}{\beta} \ln Z,
\end{align}
with $Z$ being the partition function and $\beta$ the inverse temperature. The calculation of the bounce at finite temperature is highly non-trivial because it has more intricate properties than the bounce at zero-temperature. The bounce for the transition at finite temperature depends on the Euclidean time $\tau$ and spatial radial distance $r$ and is periodic in $\tau$ direction. (For another argument of introducing the bounce with the periodic property, see Ref.\ \cite{Steingasser:2023gde}.) Finding a solution of the EoM with these features is highly non-trivial.  Nevertheless, the calculation of the bounce is crucial for understanding the transition of metastable states in a thermal bath. 

In this paper, we propose a novel method for calculating the bounce describing the transition in a thermal bath. We use the periodicity of the bounce of our interest in $\tau$ direction.  We Fourier-transform the bounce in $\tau$ direction and derive a set of EoMs for the Fourier amplitudes (which are $r$-dependent). The EoMs can be numerically solved to obtain the bounce configuration. (For other attempts to numerically calculate the bounce describing the transition in a thermal bath, see Refs.\ \cite{Ferrera:1995gs, Widyan:1998wa, Widyan:1999zg}.) Our method automatically guarantees the periodicity of the bounce in $\tau$ direction. We have applied our method of calculating the bounce for several cases, including both thin- and thick-wall cases. The validity and reliability of our method are confirmed by applying it to thin-wall case and comparing the results of numerical calculations with semi-analytic results. Additionally, we study the bounce properties in detail. For the transition in a thermal bath, there exist two types of bounces, $\tau$-dependent and $\tau$-independent bounces. We develop a procedure to calculate the critical temperature $\beta_{\rm c}^{-1}$ below which the $\tau$-independent bounce is destabilized against fluctuations and is irrelevant; such a critical temperature is obtained by the properties of the fluctuations around the $\tau$-independent bounce. According our numerical analysis, we show that the $\tau$-dependent bounce smoothly deforms into the $\tau$-independent one with the increase of the temperature in the thick-wall case while, in the thin-wall case, transition from the one to the other is discontinuous.

The remainder of this paper is organized as follows. In Section \ref{sec:transition}, based on Ref.\ \cite{Affleck:1980ac}, we provide an overview about the transition of metastable states in a thermal bath. Section \ref{sec:bounce} details our procedure of determining the bounce configuration in a thermal bath.  In Section \ref{sec:properties}, we apply our method to several models and study the properties of the bounce. Finally, Section \ref{sec:summary} summarizes the result.

\section{Transition of Metastable States in Thermal Bath}
\label{sec:transition}
\setcounter{equation}{0}
\setcounter{figure}{0}

We start with the overview of basic issues about the transition of the metastable state in a thermal bath. The transition rate can be parameterized as
\begin{align}
  \Gamma = A e^{-B},
\end{align}
where $B$ is often called as the ``bounce action'' (see below) while the prefactor $A$ takes account of the effects of quantum fluctuations. The transition rate is highly dependent on the bounce action $B$ (although the calculation $A$ is also important to precisely calculate the transition rate) and hence we focus on the precise study of $B$.

The transition rate of the metastable state in a thermal bath is related to the imaginary part of the free energy $F$.  In Ref.\ \cite{Affleck:1980ac}, the relation between ${\rm Im}F$ and $\Gamma$ has been discussed with studying a quantum system with one degree of freedom, with comparing the transition rate in quantum mechanics (adopting the semiclassical approximation) and statistical physics in a thermal bath. (In Appendix \ref{app:qm}, we summarize the argument of Ref.\ \cite{Affleck:1980ac} with its extension to the system with $N>1$ degrees of freedom.)

We apply the discussion of Ref.\ \cite{Affleck:1980ac} to the case of QFTs. We consider the QFT containing real scalar fields $\phi_a$ (with $a$ being the flavor index), whose Lagrangian density (with the scalar potential $V$) is given by
\begin{align}
  {\cal L} = \frac{1}{2} \partial_\mu \phi_a \partial^\mu \phi_a - V,
\end{align}
where the summation over the repeated flavor indices is implicit. The arguments in Appendix \ref{app:qm} can be applied to the QFT with replacing
\begin{align}
  q_i \rightarrow \phi_a (t, \vec{x}),~~~
  U \rightarrow \int d^{D-1} x 
  \left[ \frac{1}{2} (\vec{\nabla} \phi_a)(\vec{\nabla} \phi_a)
  + V
  \right].
\end{align}
Here and hereafter, the space-time dimension is implicitly $D=4$ unless otherwise stated. In addition, $\vec{\nabla}$ is the gradient operator for the $(D-1)$-dimensional spatial coordinate.

The partition function $Z\equiv{\rm Tr}\, e^{-\beta H}$ (with $H$ being the Hamiltonian) can be evaluated by using the path integral method. Regarding $e^{-\beta H}$ as an evolution operator to the periodic Euclidean-time direction (which we denote $\tau$), $Z$ is given in the following form:
\begin{align}
  Z = \int \mathcal{D} \phi e^{-S_\beta [\phi]},
\end{align}
with
\begin{align}
  S_\beta [\phi] \equiv
  \int_{-\beta/2}^{\beta/2} d\tau \int d^{D-1}x
  \left[ \frac{1}{2} (\partial_\tau \phi_a) (\partial_\tau \phi_a)
    + \frac{1}{2} (\vec{\nabla} \phi_a)(\vec{\nabla} \phi_a)
    + V \right],
\end{align}
where $\partial_\tau$ is the derivative with respect to $\tau$. In the above expression, $\phi_a$ is the variable for the path integral; it depends on $\tau$ and $(D-1)$-dimensional spatial coordinate $\vec{x}$, and obeys the periodic boundary condition $\phi_a(\tau,\vec{x})=\phi_a(\tau+\beta,\vec{x})$.

We may evaluate the partition function by the steepest descent method. For this purpose, we should derive the solution of the Euclidean equation of motion, i.e., bounce. The bounce, denoted as $\bar{\phi}$, obeys the following Euclidean EoM:
\begin{align}
  \left( \partial_\tau^2 + \vec{\nabla}^2 \right) \bar{\phi}_a
  - \left. \frac{\partial V}{\partial \phi_a} \right|_{\phi=\bar{\phi}} = 0,
\end{align}
where $\vec{\nabla}^2$ is the Laplacian in $(D-1)$-dimensional space. Assuming that the bounce is spherically symmetric with respect to the spatial coordinate, the bounce depends on two variable $\tau$ and $r$ (where $r$ is the spatial radial distance from the center of the bounce) and obeys
\begin{align}
  \left( \partial_\tau^2 + \partial_r^2 
  + \frac{D-2}{r} \partial_r \right) \bar{\phi}_a
  - \left. \frac{\partial V}{\partial \phi_a} \right|_{\phi=\bar{\phi}} = 0,
  \label{BounceEom}
\end{align}
where $\partial_r$ is the derivative with respect to $r$. We choose $(\tau,r)=(0,0)$ as the center of the bounce. Then, the bounce should satisfy the following periodic condition:
\begin{align}
  \bar{\phi}_a (\tau=-\beta/2, r) = \bar{\phi}_a (\tau=\beta/2, r),
  \label{bc_periodic}
\end{align}
as well as
\begin{align}
  \partial_\tau \bar{\phi}_a (\tau=0, r) =
  \partial_\tau \bar{\phi}_a (\tau=\beta/2, r) = 0.
  \label{bc_tau=0}
\end{align}
It also satisfies the following boundary conditions:
\begin{align}
  \partial_r \bar{\phi}_a (\tau, r=0) = 0,~~~
  \bar{\phi}_a (\tau, r\rightarrow\infty) = 0.
  \label{bc_r}
\end{align}
We implicitly assume that $\bar{\phi}_a (\tau, r)$ has $\tau$-dependence. We should note that there always exists a $\tau$-independent solution of Eq.\ \eqref{BounceEom}, satisfying the conditions \eqref{bc_periodic}, \eqref{bc_tau=0} and \eqref{bc_r}. We call such a solution as $\tau$-independent bounce $\widetilde{\phi}(r)$; it obeys
\begin{align}
  \left( \partial_r^2 
  + \frac{D-2}{r} \partial_r \right) \widetilde{\phi}_a
  - \left. \frac{\partial V}{\partial \phi_a} \right|_{\phi=\widetilde{\phi}} = 0.
  \label{EqTauindep}
\end{align}
Notice that $\widetilde{\phi}(r)$ corresponds to the ``saddle-point solution'' given in Eq.\ \eqref{hatq}.

Expanding the field around the bounce as $\phi_a=\bar{\phi}_a+\rho_a$ (with $\rho_a$ being the quantum fluctuation around the bounce),  one finds
\begin{align}
  Z \simeq Z_0 + 
  e^{-S_\beta [\bar{\phi}]}
  \int \mathcal{D}\rho
  \exp \left[ - \int_{-\beta/2}^{\beta/2} d\tau \int d^{D-1}x
    \rho_a (\tau,\vec{x}) \mathcal{M}_{ab} \rho_b (\tau,\vec{x}) \right]
  + \cdots,
  \label{pertitionfn}
\end{align}
where $Z_0$ is the zero-bounce contribution which is real, while the second term is the one-bounce contribution with $S_\beta [\bar{\phi}]$ being the bounce action.  Here, $\mathcal{M}_{ab}$ is the fluctuation operator around the bounce:
\begin{align}
  \mathcal{M}_{ab} \equiv 
  \left( -\partial_\tau^2 - \vec{\nabla}^2 \right) \delta_{ab}
  + \left. 
  \frac{\partial^2 V}{\partial\phi_a \partial\phi_b} 
  \right|_{\phi\rightarrow\bar{\phi}}.
\end{align}
The one-bounce contribution is proportional to $({\rm Det}\mathcal{M}_{ab})^{-1/2}$. Because the bounce has an unstable direction in the configuration space and $\mathcal{M}_{ab}$ has one negative eigenvalue, $({\rm Det}\mathcal{M}_{ab})^{-1/2}$ is imaginary. Consequently, $Z$ acquires an imaginary part so does $F$.  Schematically, ${\rm Im}F$ is proportional to $e^{-S_\beta [\bar{\phi}]}|{\rm Det}\mathcal{M}_{ab}|^{-1/2}$.\footnote
{There exist zero-modes in association with the translational invariance \cite{Callan:1977pt}. Extra zero-modes may also exist if the theory has an internal symmetry. In calculating the functional determinant of the fluctuation operator, these zero-modes should be properly taken care of.}

\section{Calculation of the Bounce Solution}
\label{sec:bounce}
\setcounter{equation}{0}
\setcounter{figure}{0}

In this Section, we explain our method to calculate the bounce solution which obeys Eqs.\ \eqref{BounceEom}, \eqref{bc_periodic}, \eqref{bc_tau=0} and \eqref{bc_r}. We consider the scalar potential of the following form:
\begin{align}
  V = g^{(1)}_a \phi_a 
  + \frac{1}{2} g^{(2)}_{ab} \phi_a \phi_b
  + \frac{1}{3} g^{(3)}_{abc} \phi_a \phi_b \phi_c
  + \frac{1}{4} g^{(4)}_{abcd} \phi_a \phi_b \phi_c \phi_d,
\end{align}
where $g^{(1)}_a$, $g^{(2)}_{ab}$, $g^{(3)}_{abc}$ and $g^{(4)}_{abcd}$ are constants
and are invariant under permutations of the flavor indices. Summation over the repeated flavor indices is implicit.

For the calculation of the bounce, we use the fact that the bounce of our interest is periodic in $\tau$ direction with the periodicity of $\beta$.  With the use of Eq.\ \eqref{bc_tau=0}, we expand the solution as
\begin{align}
  \bar{\phi}_a (\tau, r) = 
  \sum_{n=0}^\infty \bar{\varphi}_{a,n} (r)
  \cos \left( \frac{2n\pi}{\beta} \tau \right).
\end{align}
Then, we can find
\begin{align}
  \left[ \partial_r^2 + \frac{D-2}{r} \partial_r
    - \left( \frac{2 n \pi}{\beta} \right)^2 \right] \bar{\varphi}_{a,n} = &\,
  g^{(1)}_a \delta_{n,0}
  + g^{(2)}_{ab} \bar{\varphi}_{b,n}
  \nonumber \\ &\,
  + g^{(3)}_{abc} \sum_{\ell_1,\ell_2} C_{n,\ell_1,\ell_2,0} \bar{\varphi}_{b,\ell_1} \bar{\varphi}_{c,\ell_2}
  \nonumber \\ &\,
  + g^{(4)}_{abcd} \sum_{\ell_1,\ell_2,\ell_3} C_{n,\ell_1,\ell_2,\ell_3} \bar{\varphi}_{b,\ell_1} \bar{\varphi}_{c,\ell_2}\bar{\varphi}_{d,\ell_3},
  \label{EqForNumericalCalc}
\end{align}
where 
\begin{align}
  C_{n,\ell_1,\ell_2,\ell_3} = \frac{1}{4(1+\delta_{n,0})}
  \sum_{\sigma_1=\pm}
  \sum_{\sigma_2=\pm}
  \sum_{\sigma_3=\pm}
  \delta_{n+\sigma_1 \ell_1+\sigma_2 \ell_2+ \sigma_3 \ell_3,0}.
  \label{C_nlll}
\end{align}

The bounce solution is obtained by numerically solving Eq.\ \eqref{EqForNumericalCalc}; we can use, for example, the Newton's method to solve Eq.\ \eqref{EqForNumericalCalc} with discretizing the $r$ direction. More detail about the numerical calculation is discussed in Appendix \ref{app:algorism}. We also note here that the method described above can be applied to obtain the $\tau$-independent bounce $\tilde{\phi}_a$ with performing our method with taking $\bar{\varphi}_{a,n}=0$ for $n\neq 0$.

\section{Studying the Properties of the Bounce}
\label{sec:properties}
\setcounter{equation}{0}
\setcounter{figure}{0}

In this section, we demonstrate that our procedure of calculating the bounce really works by numerically solving the equation given in the previous section. Because of the limitation of our computational power, we consider QFT models with one real scalar field.  We parameterize the scalar potential as
\begin{align}
  V (\phi) = V_0 (\phi) + V_\epsilon (\phi),
  \label{Vphi}
\end{align}
where
\begin{align}
  V_0 (\phi) = &\, \lambda \phi^2 (\phi - v)^2,
  \\
  V_\epsilon (\phi) = &\, \epsilon \lambda v (2\phi^3 - 3v\phi^2).
\end{align}
Here, $\lambda$ and $v$ are positive constants while $0<\epsilon<\frac{1}{3}$. The potential has the false vacuum at $\phi=0$ (with $V(0)=0$) and the true vacuum at $\phi=v$ (with $V(v)=-\epsilon\lambda v^4$).  The peak of the potential is at $\phi=\frac{1-3\epsilon}{2}v$.  We also define the ``scalar mass'' parameter around the false vacuum as
\begin{align}
  m_\phi \equiv \left. \sqrt{\frac{\partial^2 V}{\partial \phi^2}}
  \right|_{\phi=0} = \sqrt{2\lambda (1-3\epsilon)}v.
\end{align}

For the potential given in Eq.\ \eqref{Vphi}, we determine the bounce solution by using the method introduced in the previous section. We also calculate the bounce action, denoted as $\bar{S}$, as
\begin{align}
  \bar{S} \equiv \Omega_{D-1} 
  \int_{-\beta/2}^{\beta/2} d\tau \int_0^\infty dr r^{D-2}
  \left[ \bar{\phi} \left( -\partial_\tau^2 - \partial_r^2 - \frac{D-2}{r} \partial_r \right) \bar{\phi} + V \right].
\end{align}
Here, $\Omega_d$ denotes the surface area of the $d$-dimensional unit sphere ($\Omega_4=2\pi^2$ and $\Omega_3=4\pi$).  Embedding the $\tau$-independent solution into the space with periodic $\tau$ direction, we also define the action of such an object as
\begin{align}
  \widetilde{S} \equiv \beta \Omega_{D-1} 
  \int_0^\infty dr r^{D-2}
  \left[ \widetilde{\phi} \left( - \partial_r^2 - \frac{D-2}{r} \partial_r \right) \widetilde{\phi} + V \right].
\end{align}
Notice that $\widetilde{S}$ is proportional to $\beta$. We also note here that $\tilde{S}/\beta$ is the action of $O(D-1)$ symmetric bounce in $(D-1)$-dimensional Euclidean space which determines the transition rate in the high temperature limit \cite{Affleck:1980ac, Linde:1980tt, Linde:1981zj}.

\subsection{Thin-wall case}

We first consider the so-called thin-wall case, which is realized when $\epsilon\ll 1$.  In such a case, the difference of the potential heights of the false and true vacua becomes much smaller than the potential height at the peak.

\begin{table}[t]
  \begin{center}
    \begin{tabular}{l|rrr}
      \hline\hline
      & Model 1 & Model 2 & Model 3
      \\
      \hline
      $v$ & 1 & 1 & 1
      \\
      $\lambda$ & $2.50\times 10^{-1}$ & $2.50\times 10^{-1}$&$2.50\times 10^{-1}$
      \\
      $\epsilon$ & $2.50\times 10^{-2}$ & $5.00\times 10^{-2}$& $1.00\times 10^{-1}$
      \\
      $\bar{S}$: Numerical & $1.05\times 10^{5}$& $1.29\times 10^{4}$& $1.52\times 10^{3}$
      \\
      $\bar{S}$: {\tt CosmoTransitions} & $1.05\times 10^{5}$&$1.29\times 10^{4}$& $1.52\times 10^{3}$
      \\
      $\bar{S}$: Eq.\ \eqref{Sbar(approx)} &$1.05\times 10^{5}$&$1.32\times 10^{4}$& $1.64\times 10^{3}$
      \\
      $\bar{r}$: Numerical &$5.65\times 10^1$&$2.81\times 10^1$&$1.38\times 10^1$
      \\
      $\bar{r}$: Eq.\ \eqref{rbar(approx)} & $5.66\times 10^1$&$2.83\times 10^1$&$1.41\times 10^1$
      \\
      $\beta_c$: Numerical &$1.67\times10^{2}$&$8.35\times 10^1$&$4.12\times 10^1$
      \\
      $\beta_c$: Eq.\ \eqref{betac(approx)} & $1.67\times10^{2}$&$8.31\times 10^1$&$4.07\times 10^1$
      \\
      \hline\hline
    \end{tabular}
    \caption{Model parameters used in our numerical analysis for the thin-wall case. We also show the results of numerical estimations of $\bar{S}$ and $\bar{r}$ in the low-temperature limit, and $\beta_c$; here $\bar{S}$ and $\bar{r}$ are estimated at $\beta=3\bar{r}$.  The space-time dimension is $D=4$.  We take the unit such that $v=1$. Approximated values of $\bar{S}$, $\bar{r}$ and $\beta_c$ are also shown, which are based on Eqs.\ \eqref{Sbar(approx)}, \eqref{rbar(approx)} and \eqref{betac(approx)}, respectively.}
    \label{table:results}
  \end{center}
\end{table}

Using the procedure introduced in Section \ref{sec:bounce}, we calculate the bounce solution $\bar{\phi}$ for several cases. In the present analysis, we consider the cases of $\epsilon = 0.025$, $0.05$, and $0.1$; model parameters used in our analysis are summarized in Table \ref{table:results}.  We also numerically estimate several physical quantities, i.e., $\bar{S}$ and the bounce size $\bar{r}$ in the low-temperature limit, as well as the critical inverse temperature $\beta_{\rm c}$ below which the $\tau$-independent bounce becomes destabilized against fluctuations (see the following discussion). Results are also shown in Table \ref{table:results}.

\begin{figure}
    \centering
    \includegraphics[width=1\textwidth]{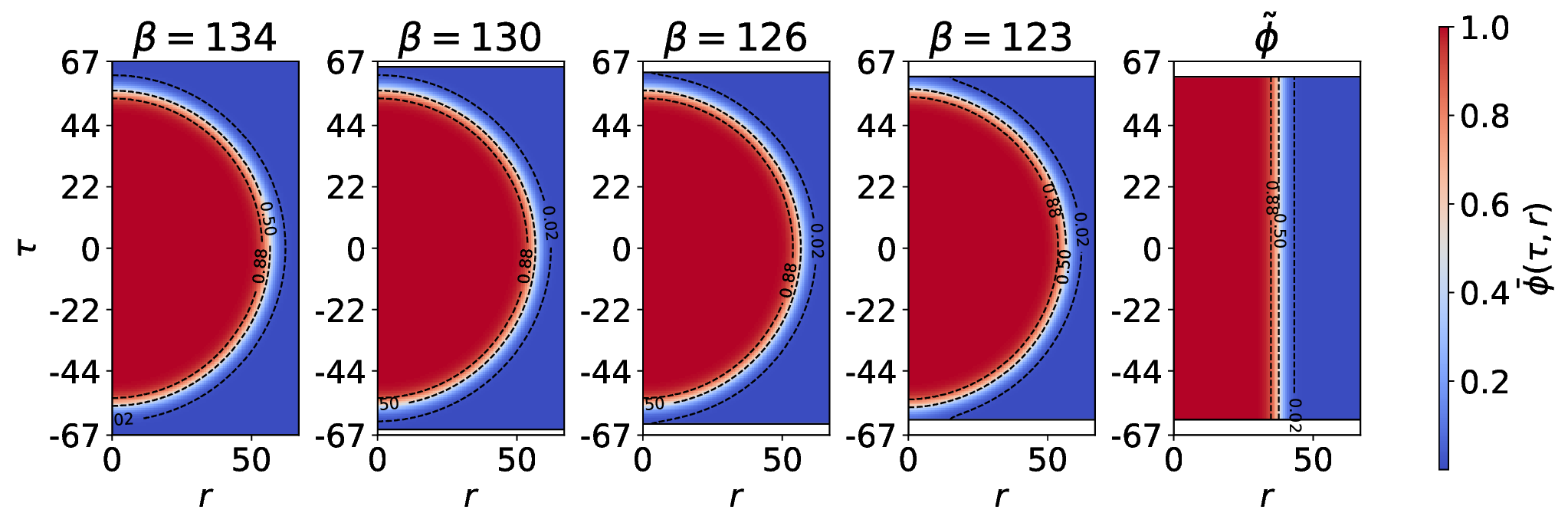}
    \includegraphics[width=1\textwidth]{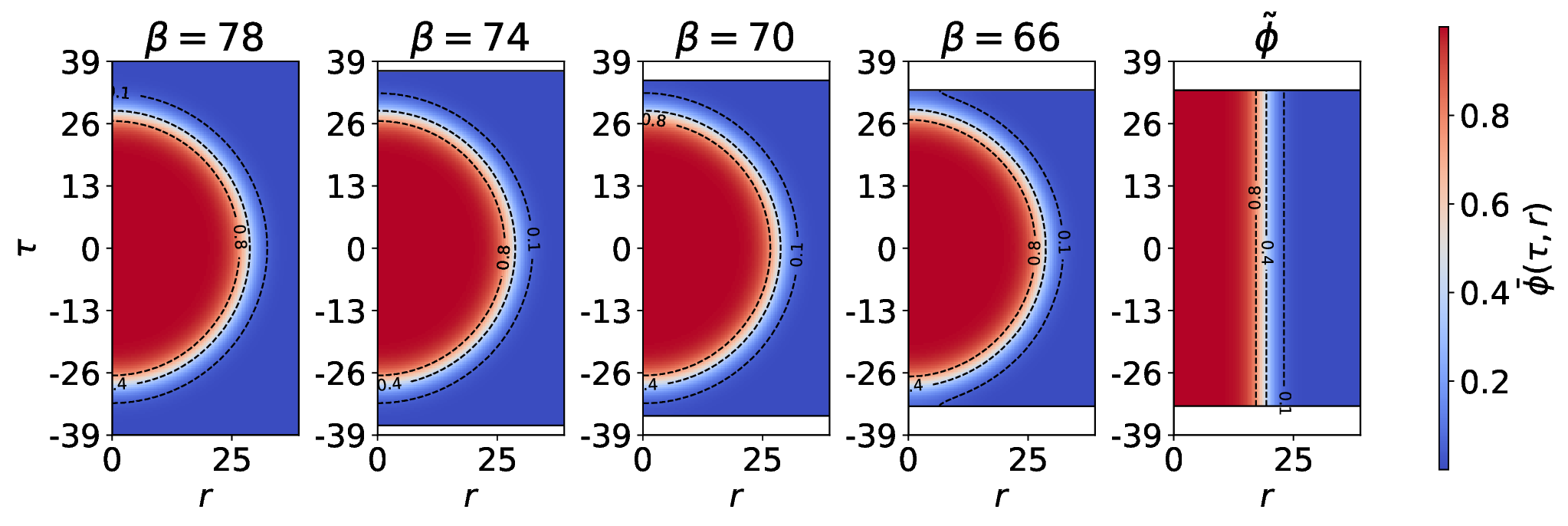}
    \includegraphics[width=1\textwidth]{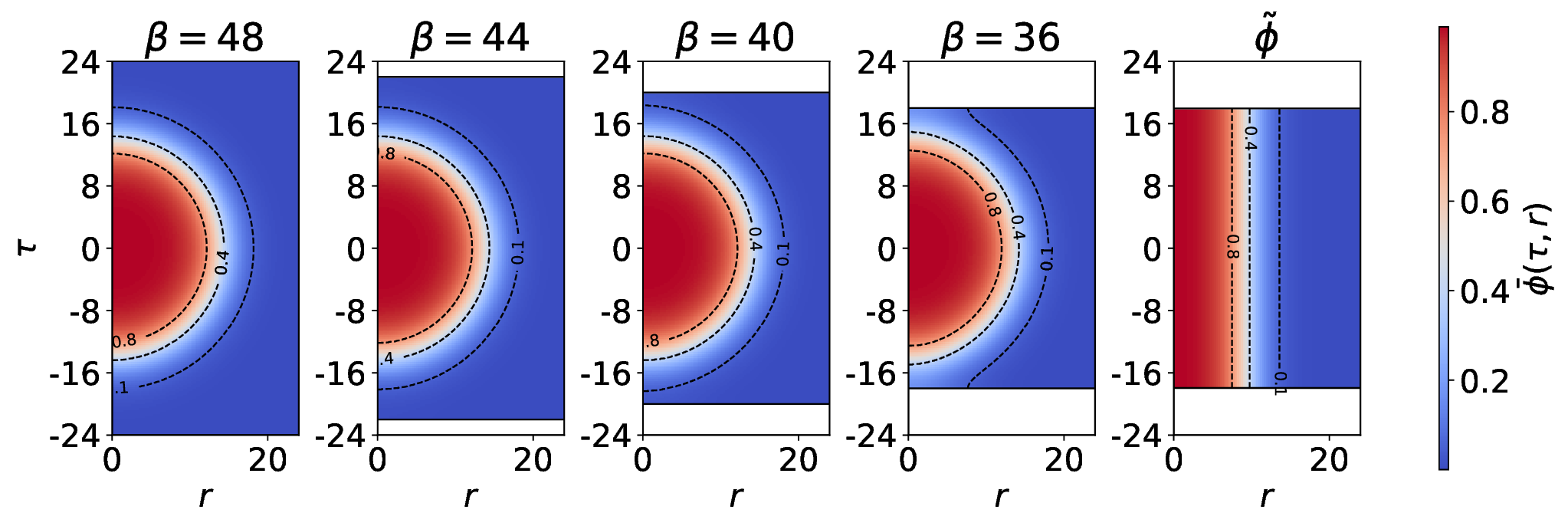}
    \caption{Bounce configuration for the thin-wall cases: Model 1 with $\epsilon = 0.025$ (top), Model 2 with $\epsilon = 0.05$ (middle), and Model 3 with $\epsilon = 0.1$ (bottom). Color intensity represents the field value; the false and true vacua are $\phi=0$ and $1$, respectively. The figures with ``$\tilde{\phi}$'' on the top show the $\tau$-independent bounce, while others are $\tau$-dependent ones for different temperatures. Lines in the figures indicate contours of constant $\bar{\phi}$ or $\tilde{\phi}$.}
    \label{fig:heatmapthin}
\end{figure}

In Fig.\ \ref{fig:heatmapthin}, we show the bounce configuration for Models 1, 2 and 3; both $\tau$-dependent bounce (for several different temperatures) and $\tau$-independent bounce are shown. In all the cases, we can see that the thickness of the wall is much narrower than the bounce size so that the thin-wall approximation is applicable. In the thin-wall case, with the increase of the temperature, the $\tau$-dependent bounce almost keeps its spherical shape and that it disappears at a certain temperature (which we call $\beta^{-1}_*$).  For Models 1, 2, and 3 (with $\epsilon = 0.025$, $0.05$, and $0.1$, respectively), we found $\beta_*\simeq 2.2\bar{r}$, $2.3\bar{r}$ and $2.5\bar{r}$, respectively.

In the thin-wall case, we can have insights into the bounce properties with semi-analytic approach \cite{Coleman:1977py, Garriga:1994ut}. Particularly, at low enough temperature at which the bounce size is much smaller than $\beta$, $\bar{\phi}$ is well approximated by the bounce at zero temperature, which is given as the $O(D)$ symmetric bounce in $D$-dimensional Euclidean space. In such a case, the bounce approximately obeys the following equation:
\begin{align}
  \left( \partial_s^2 
  + \frac{D-1}{s} \partial_s \right) \bar{\phi}
  - \left. \frac{\partial V}{\partial \phi} \right|_{\phi=\bar{\phi}} \simeq 0,
  \label{O(D)bounce}
\end{align}
where $s\equiv\sqrt{\tau^2+r^2}$ is the radius parameter in $D$-dimensional Euclidean space.  In the thin-wall case, we may approximate the potential as a double-well one (with taking $\epsilon\rightarrow 0$) with neglecting the friction term (i.e., second term in the parenthesis of Eq.\ \eqref{O(D)bounce}) to obtain the following approximate form:
\begin{align}
  \bar{\phi} (\tau,r) \simeq \frac{1}{2} 
  \left[ 1 - \tanh \left\{ \frac{1}{2} m_\phi (s - \bar{r}) \right\} \right]
  v.
  \label{phibar(approx)}
\end{align}
The ``bounce size'' $\bar{r}$ is estimated by taking account of the effects of $V_\epsilon$ as well as the friction term in Eq.\ \eqref{O(D)bounce}:
\begin{align}
    \bar{r} 
    \simeq \frac{D-1}{3\epsilon} m_\phi^{-1},
    \label{rbar(approx)}
\end{align}
where, in our numerical calculation, $m_\phi$ is evaluated with taking $\epsilon\rightarrow 0$ limit in estimating $\bar{r}$. Notice that, in the present case, the bounce size $\bar{r}$ is much larger than the width of the wall, which is $\sim m_\phi^{-1}$, because $\epsilon\ll 1$.  Using Eq.\ \eqref{phibar(approx)}, the bounce action is estimated as
\begin{align}
  \bar{S} \simeq \frac{\Omega_D}{6D} m_\phi v^2 \bar{r}^{D-1}.
  \label{Sbar(approx)}
\end{align}
We have checked the validity of these approximated values. Based on the argument above, we expect that the properties of the $\tau$-dependent bounce becomes insensitive to $\beta$ if $\beta$ is sufficiently larger than $\bar{r}$.  We have numerically derived $\bar{\phi}$ for $\beta=3\bar{r}_{\eqref{rbar(approx)}}$ (with $\bar{r}_{\eqref{rbar(approx)}}$ being the bounce size evaluated by using Eq.\ \eqref{rbar(approx)}) and estimated the bounce size by solving $\bar{\phi}(\tau=0,\bar{r})=\frac{1}{2}\bar{\phi}(\tau=0,r=0)$. In addition, $\bar{\phi}$ is also used to evaluate the bounce action $\bar{S}$.  The results, as well as $\bar{r}$ estimated based on Eq.\ \eqref{rbar(approx)} and $\bar{S}$ from Eq.\ \eqref{Sbar(approx)}, are shown in Table \ref{table:results}. We can see that Eqs.\ \eqref{rbar(approx)} and \eqref{Sbar(approx)} well approximate $\bar{r}$ and $\bar{S}$, respectively. In order to check the validity of our result, we also estimate the action of the $\tau$-dependent bounce in the low-temperature limit by using the public code {\tt CosmoTransitions} \cite{Wainwright:2011kj} as well as by the gradient-flow method proposed in Ref.\ \cite{Chigusa:2019wxb}; in such calculations, the bounce in the low-temperature limit is approximated as an $O(D)$ symmetric object. We have checked that the numerical results based on our procedure well agree with those based on the other methods; the $\tau$-dependent bounce action in the low-temperature limit calculated by {\tt CosmoTransitions} is also shown in Table \ref{table:results}.

As in the case of $\tau$-dependent bounce in the low-temperature limit, the $\tau$-independent bounce $\widetilde{\phi}$ can be approximated as
\begin{align}
  \widetilde{\phi} (r) \simeq \frac{1}{2} 
  \left[ 1 - \tanh \left\{ \frac{1}{2} m_\phi (r - \widetilde{r}) \right\} \right]
  v,
\end{align}
with
\begin{align}
  \widetilde{r} \simeq \frac{D-2}{3\epsilon} m_\phi^{-1}.
  \label{rtilde(approx)}
\end{align}
Then, we find
\begin{align}
  \widetilde{S} \simeq \frac{\Omega_{D-1}}{6(D-1)} m_\phi v^2 \beta\widetilde{r}^{D-2}.
  \label{Stilde(approx)}
\end{align}
Using Eqs.\ \eqref{Sbar(approx)} and \eqref{Stilde(approx)}, we can estimate $\beta_{\rm eq}$, the inverse temperature at which the actions of $\tau$-dependent and $\tau$-independent bounces become equal:
\begin{align}
  \beta_{\rm eq} \simeq \frac{\Omega_D}{\Omega_{D-1}} 
  \frac{(D-1)^{D-1}}{D(D-2)^{D-2}} \bar{r}.
  \label{beta_eq}
\end{align}

\begin{figure}
    \centering
    \includegraphics[width=9.5cm]{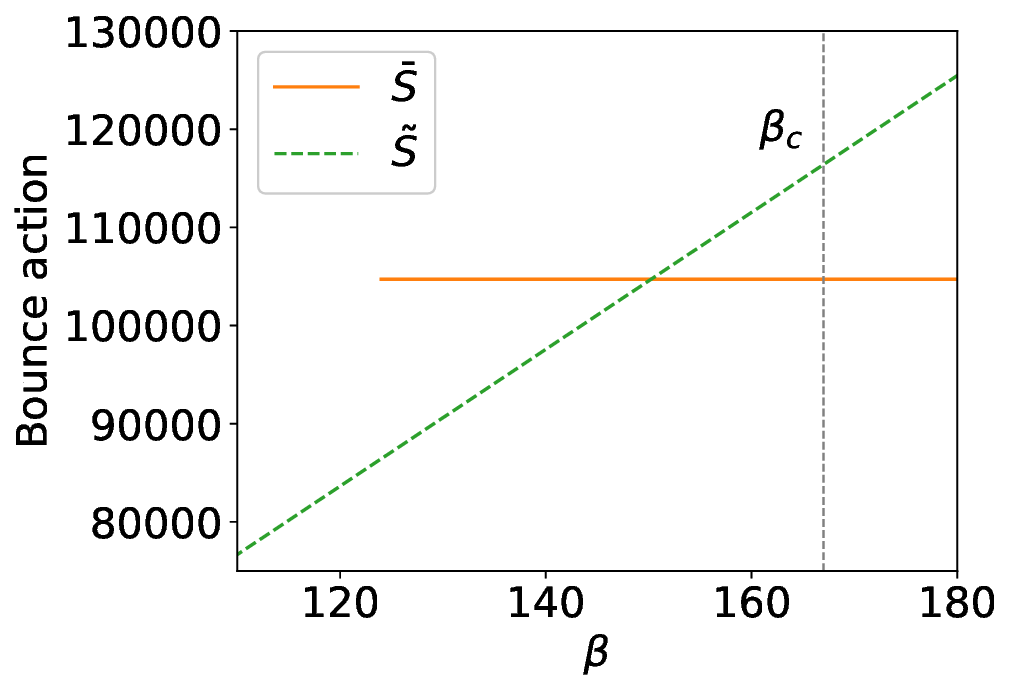}
    \includegraphics[width=8.5cm]{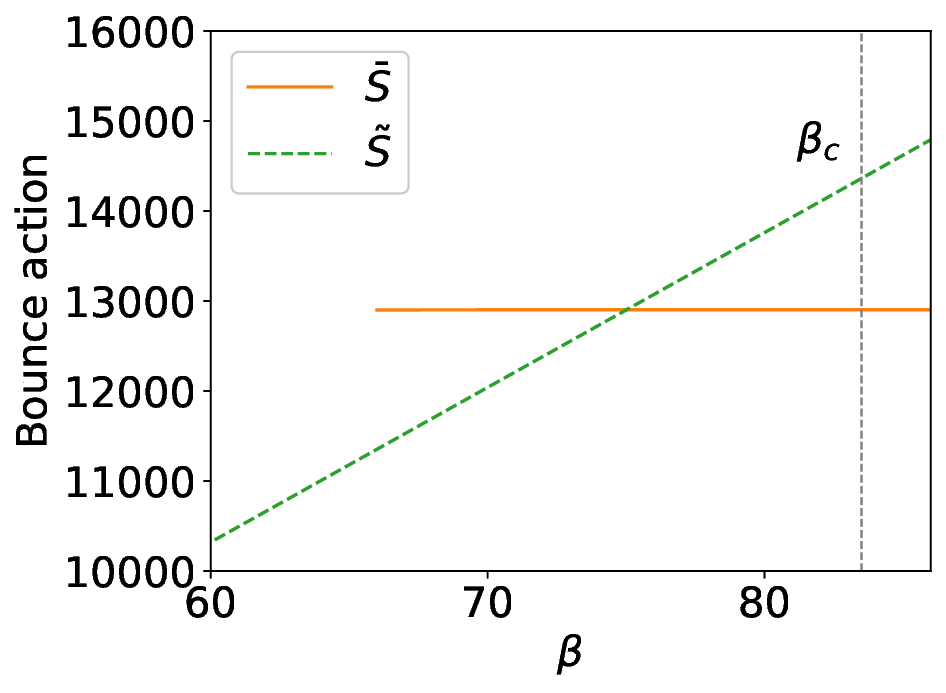}
    \includegraphics[width=8.5cm]{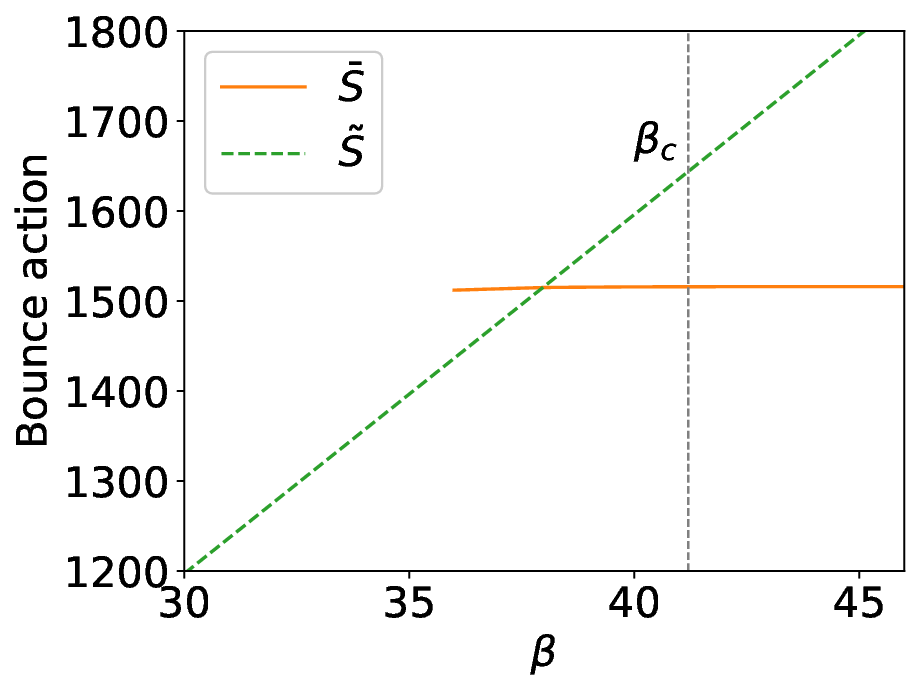}
    \caption{$\bar{S}$ and $\tilde{S}$ as functions of $\beta$ for Model 1 with $\epsilon = 0.025$ (top), Model 2 with $\epsilon = 0.05$ (middle), and Model 3 with $\epsilon = 0.1$ (bottom). The vertical dashed line shows $\beta_{\rm c}$.}
    \label{fig:Sethin}
\end{figure}

Now, we discuss the temperature dependences of bounce actions. In Fig.\ \ref{fig:Sethin}, we plot $\bar{S}$ and $\widetilde{S}$ as functions of $\beta$ for Models 1, 2, and 3 (with $\epsilon = 0.025$, $0.05$, and $0.1$, respectively). As expected, the action of the $\tau$-dependent bounce $\bar{S}$ is almost independent of the temperature (as far as $\beta\gtrsim\beta_*$) while the action of the $\tau$-independent bounce $\tilde{S}$ is proportional to $\beta$.  The line of $\bar{S}$ is terminated at $\beta=\beta_*$ because the $\tau$-dependent bounce does not exist for $\beta<\beta_*$.  When $\beta\sim\beta_*$, the action of the $\tau$-dependent bounce is larger than that of the $\tau$-independent bounce \cite{Garriga:1994ut}. We expect that the path integral for the transition is dominated by the field configuration with smaller action and that the temperature-dependence of the transition rate changes discontinuously at $\beta\sim\beta_*$.

One important to understand the property of the $\tau$-independent bounce $\widetilde{\phi}$ is the negative eigenvalue of the fluctuation operator around it.  According to the discussion in Appendix \ref{app:qm}, the $\tau$-independent bounce is expected to become irrelevant when $\beta>2\pi/\mu$, with $-\mu^2$ being the negative eigenvalue of the fluctuation operator around $\widetilde{\phi}$. This fact may be reinterpreted in QFTs as follows. In order to perform the path integral around $\widetilde{\phi}$, let us expand the scalar field around $\widetilde{\phi}$ as 
\begin{align}
  \phi (\tau, r) = \widetilde{\phi} (r) + \rho (\tau,r),
\end{align}
where $\rho$ is the quantum fluctuation to be integrated out.  Using the periodicity of $\phi$, $\rho$ can be expanded as
\begin{align}
  \rho (\tau, r) = 
  \sum_{n=-\infty}^\infty \rho_{n} (r)
  \exp \left( i \frac{2n\pi}{\beta} \tau \right),
\end{align}
with $\rho_{-n}^*=\rho_n$. Adopting the steepest descent method to calculate the contribution of the $\tau$-independent bounce to the partition function (which we denote as $\widetilde{Z}_1$), we expect
\begin{align}
  \widetilde{Z}_1 \simeq
  e^{-\widetilde{S}} \prod_{n=-\infty}^\infty \int \mathcal{D}\rho_n
  \exp \left[ - \Omega_{D-2} \beta
    \int_0^\infty dr r^{D-2} 
    \rho_n (r) \widetilde{\mathcal{M}}_n \rho_n (r) \right],
\end{align}
where
\begin{align}
  \widetilde{\mathcal{M}}_n \equiv -\partial_r^2 - \frac{D-2}{r} \partial_r
  + \left. \frac{\partial^2 V}{\partial \phi^2} 
  \right|_{\phi\rightarrow\widetilde{\phi}}
  + \left( \frac{2n\pi}{\beta} \right)^2,
\end{align}
which implies
\begin{align}
  \widetilde{Z}_1 \propto
  e^{-\widetilde{S}}
  \prod_{n=-\infty}^\infty \left( \mbox{Det} \widetilde{\mathcal{M}}_n \right)^{-1/2}.
\end{align}
Notice that $\widetilde{\mathcal{M}}_0$ is the fluctuation operator for the false vacuum decay at zero-temperature in $(D-1)$-dimensional space-time, which has one negative eigenvalue (denoted as $-\mu^2$). The path integral to the direction of the eigenstate in association with the negative eigenvalue of $\widetilde{\mathcal{M}}_0$ can be performed with a proper analytic continuation; consequently, the free energy acquires an imaginary part \cite{Coleman:1977py, Callan:1977pt, Affleck:1980ac}.  However, at low temperature, other fluctuation operators may also have negative eigenvalues. This is because an eigenvalue of $\widetilde{\mathcal{M}}_{0}$ and that of $\widetilde{\mathcal{M}}_{n}$, denoted as $\xi_0$ and $\xi_n$, respectively, are related to each other as $\xi_n=\xi_0+(2n\pi\beta^{-1})^2$, implying that the smallest eigenvalue of $\widetilde{\mathcal{M}}_{n}$ is $-\mu^2+(2n\pi\beta^{-1})^2$. Thus, if the temperature $\beta^{-1}$ is low enough, some of $\widetilde{\mathcal{M}}_{n}$ with $n\neq 0$ have negative eigenvalues.  In particular, $\widetilde{\mathcal{M}}_{1}$ and $\widetilde{\mathcal{M}}_{-1}$ have negative eigenvalues when $\beta>\beta_{\rm c}$, where
\begin{align}
  \beta_{\rm c} \equiv \frac{2\pi}{\mu}.
\end{align}
We can see that $\beta_{\rm c}^{-1}$ is the temperature at which the $\tau$-independent bounce becomes irrelevant according to the quantum-mechanical argument given in Ref.\ \cite{Affleck:1980ac} (see also Appendix \ref{app:qm}). We expect that, at $\beta>\beta_{\rm c}$, the $\tau$-independent bounce $\widetilde{\phi}$ is unstable against fluctuations and does not contribute to ${\rm Im}F$ in QFTs.

The negative eigenvalue of $\widetilde{\mathcal{M}}_0$ can be evaluated by using the procedure discussed in Refs.\ \cite{Gelfand:1959nq, Dashen:1974ci, Coleman:1985rnk, Kirsten:2004qv, Kirsten:2003py, Endo:2017tsz}. Consider the solution of the following differential equation:
\begin{align}
  \left( \widetilde{\mathcal{M}}_0 - \xi \right) f(\xi; r) = 0,
  \label{eigenvalueeq}
\end{align}
with $\partial_r f(\xi;r=0) = 0$, where $\xi$ is a real parameter.  Then, when $\xi$ is equal to one of the eigenvalues of $\widetilde{\mathcal{M}}_0$, $f(\xi;r\rightarrow\infty)=0$. We can find the eigenvalues of $\widetilde{\mathcal{M}}_0$ with studying the $\xi$-dependence of $f(\xi;r\rightarrow\infty)$; such an analysis can be performed numerically.  In particular, the negative eigenvalue can be found with scanning the $\xi$ within the region of $\xi<0$. (Notice that $\widetilde{\mathcal{M}}_0$ has only one negative eigenvalue.)  We numerically solve Eq.\ \eqref{eigenvalueeq} using $\tilde{\phi}$ obtained by our method and estimate $\mu$, based on which $\beta_{\rm c}$ is calculated. The result is shown in Fig.\ \ref{fig:Sethin} for each model (vertical dashed line in the figure) as well as in Table \ref{table:results}. According to our numerical result, in the thin-wall case, $\beta_*<\beta_{\rm eq}<\beta_{\rm c}$ is indicated; such a relation was obtained in Ref.\ \cite{Garriga:1994ut} based on semi-analytic approximation.  For $\beta_*<\beta<\beta_{\rm c}$, $\tau$-dependent and $\tau$-independent bounces coexist.

For the thin-wall case, we may also find an approximated formula of $\beta_{\rm c}$ using the following relation:
\begin{align}
  \widetilde{\mathcal{M}}_0 \partial_r \widetilde{\phi} =
  -\frac{D-2}{r^2} \partial_r \widetilde{\phi},
  \label{M0dphihat}
\end{align}
which is obtained by differentiating Eq.\ \eqref{EqTauindep} with respect to $r$. In the thin-wall limit, $\partial_r\widetilde{\phi}$ is almost zero except $r\sim \widetilde{r}$. Then, in the right-hand side of Eq.\ \eqref{M0dphihat}, we may approximately replace as $\frac{D-2}{r^2}\rightarrow\frac{D-2}{\widetilde{r}^2}$ and regard $\partial_r \widetilde{\phi}$ as the eigenfunction in association with the negative eigenvalue of $\widetilde{\mathcal{M}}_0$.  Then,
\begin{align}
  \mu^2 \simeq \frac{D-2}{\widetilde{r}^2},
\end{align}
based on which we obtain \cite{Garriga:1994ut}
\begin{align}
  \beta_{\rm c} \simeq \frac{2 \pi}{\sqrt{D-2}} \widetilde{r}
  \simeq \frac{2 \pi\sqrt{D-2}}{D-1} \bar{r}.
  \label{betac(approx)}
\end{align}
The value of $\beta_{\rm c}$ based on this approximation is also shown in Table \ref{table:results}; we can see that the above expression well reproduces the numerical result.

\subsection{Thick-wall case}

Next we consider the thick-wall case where $\epsilon$ is sizable. As examples, we consider the scalar potential \eqref{Vphi} with $\epsilon=0.15$, $\epsilon=0.2$ and $\lambda=0.25$ (Models 4, 5 and 6). The Model parameters are summarized in Table \ref{table:thickwall}. (As we will see below, Model 4 may not be a good example of the case of thick-wall limit, but we consider Model 4 to show how the bounce changes from the thin-wall case to the thick-wall case.) For those models, we calculate the bounce solution with our method for various temperatures, based on which the bounce action $\bar{S}$ is obtained. The values of $\bar{S}$ in the low-temperature limit and $\beta_{\rm c}$ can be found in Table \ref{table:thickwall}. For comparison, we also show $\bar{S}$ in the low-temperature limit obtained by {\tt CosmoTransitions} and that by the semi-analytic formula relevant for the thin-wall case (see Eq.\ \eqref{Sbar(approx)}) in the table.

\begin{table}[t]
  \begin{center}
    \begin{tabular}{l|rrr}
      \hline\hline
      & Model 4 & Model 5 & Model 6
      \\
      \hline
      $v$ & 1 & 1 & 1
      \\
      $\lambda$ & $2.50\times 10^{-1}$ & $2.50\times 10^{-1}$&$2.50\times 10^{-1}$
      \\
      $\epsilon$ & $1.50\times 10^{-1}$ & $2.00\times 10^{-1}$& $2.50\times 10^{-1}$
      \\
      $\bar{S}$: Numerical & $4.00\times 10^2$ & $1.39\times 10^2$& $5.00\times 10^1$
      \\
      $\bar{S}$: {\tt CosmoTransitions} & $4.02\times 10^2$& $1.38\times 10^2$& $4.99\times 10^1$
      \\
      $\bar{S}$: Eq.\ \eqref{Sbar(approx)} & $4.87\times 10^2$& $2.06\times 10^2$& $1.05\times 10^2$
      \\
      $\beta_c$: Numerical & $2.67\times 10^1$ &$1.99\times 10^1$& $1.74\times 10^1$
      \\
      \hline\hline
    \end{tabular}
    \caption{Model parameters used in our numerical analysis for the thick-wall case. The space-time dimension is $D=4$. We also show $\bar{S}$ in the low-temperature limit and $\beta_{\rm c}$.}
    \label{table:thickwall}
  \end{center}
\end{table}

From the calculation of the bounce action $\bar{S}$, we can see that our method can perform a reliable calculation of the bounce at finite temperature even in the thick-wall case. Our numerical results on $\bar{S}$ well agree with those obtained by using {\tt CosmoTransitions} (see Table \ref{table:thickwall}).  We can also see that, for Models 4, 5, and 6, Eq.\ \eqref{Sbar(approx)}, which is based on the analysis in the thin-wall limit, does not give a good approximation as expected.

\begin{figure}
    \centering
    \includegraphics[width=1\textwidth]{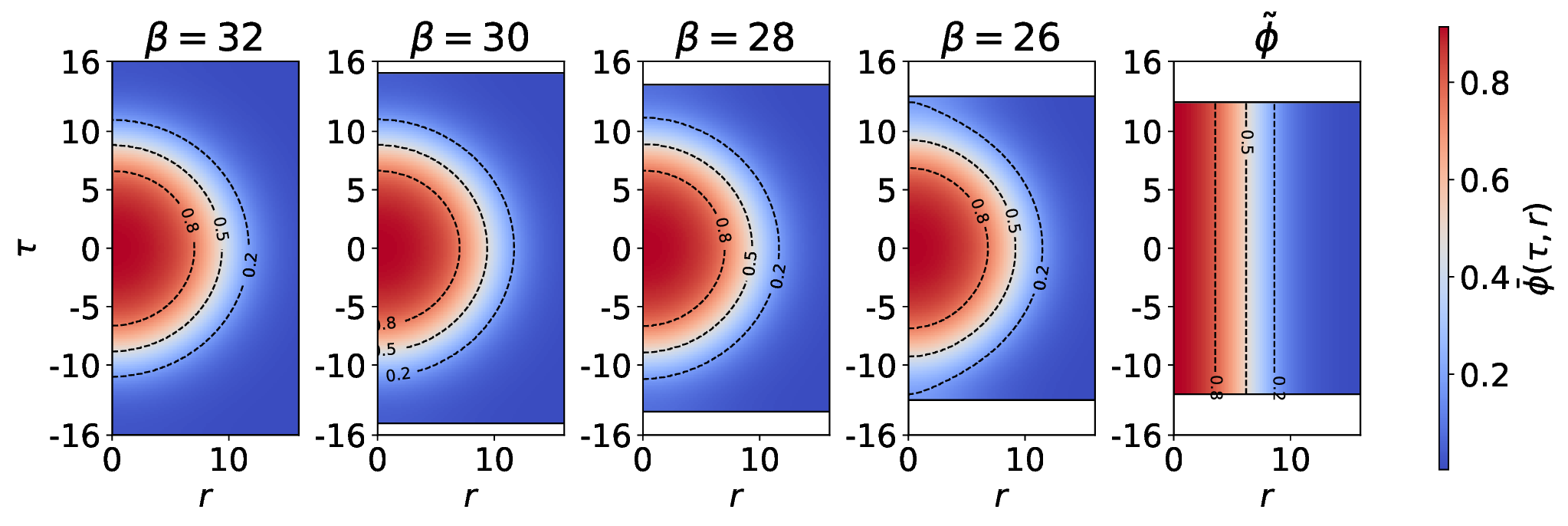}
    \includegraphics[width=1\textwidth]{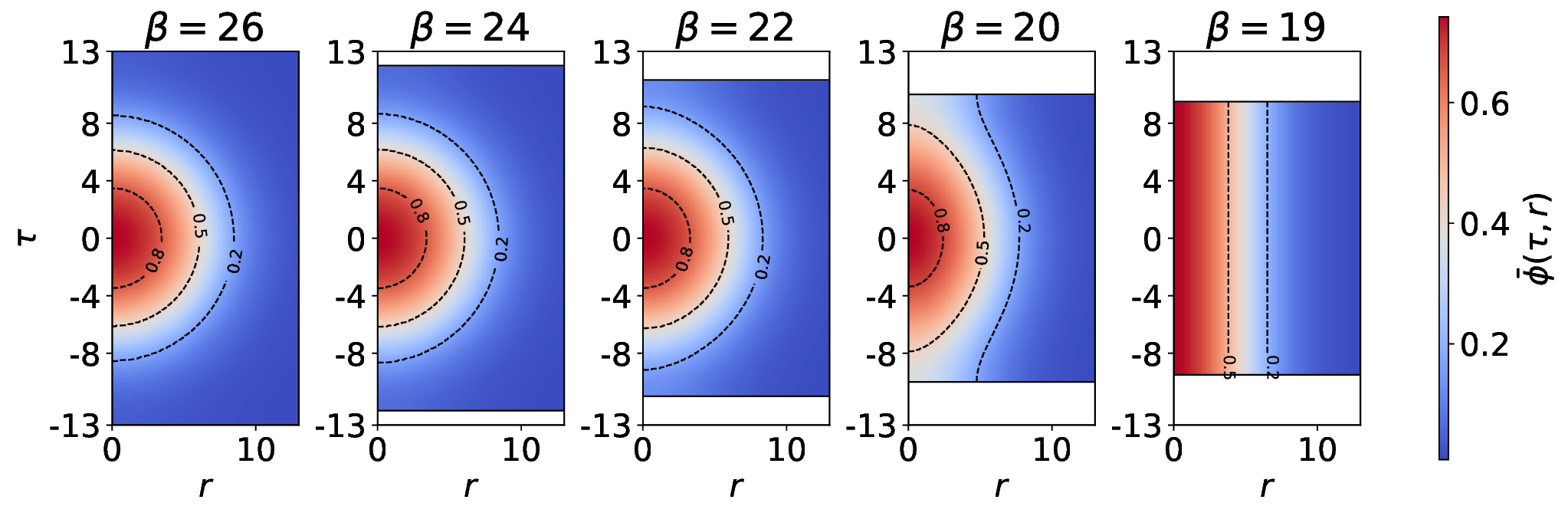}
    \includegraphics[width=1\textwidth]{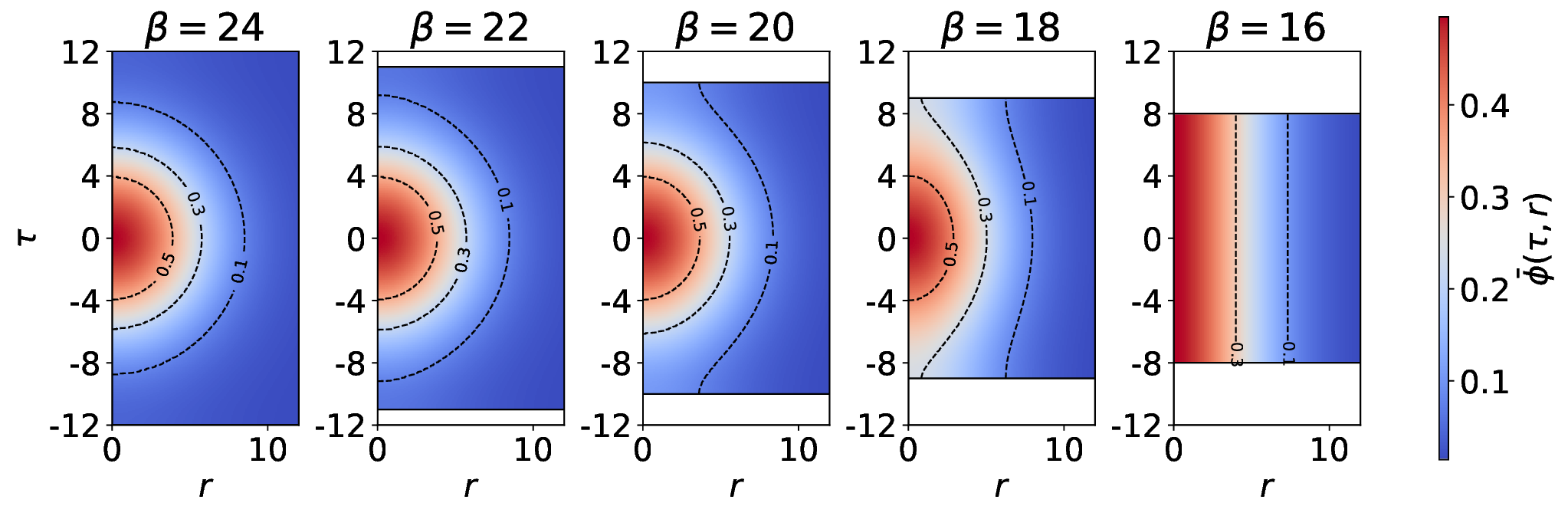}
    \caption{Bounce configuration for the thick-wall cases: Model 4 with $\epsilon=0.15$ (top), Model 5 with $\epsilon=0.2$ (middle) and Model 6 with $\epsilon=0.25$ (bottom). The figure with ``$\tilde{\phi}$'' on the top shows the $\tau$-independent bounce, while others are $\tau$-dependent ones for different temperatures. Lines in the figures indicate contours of constant $\bar{\phi}$ or $\tilde{\phi}$.}
    \label{fig:heatmapthick}
\end{figure}

\begin{figure}
    \centering
    \begin{minipage}{\linewidth}
        \centering
        \includegraphics[width=8cm]{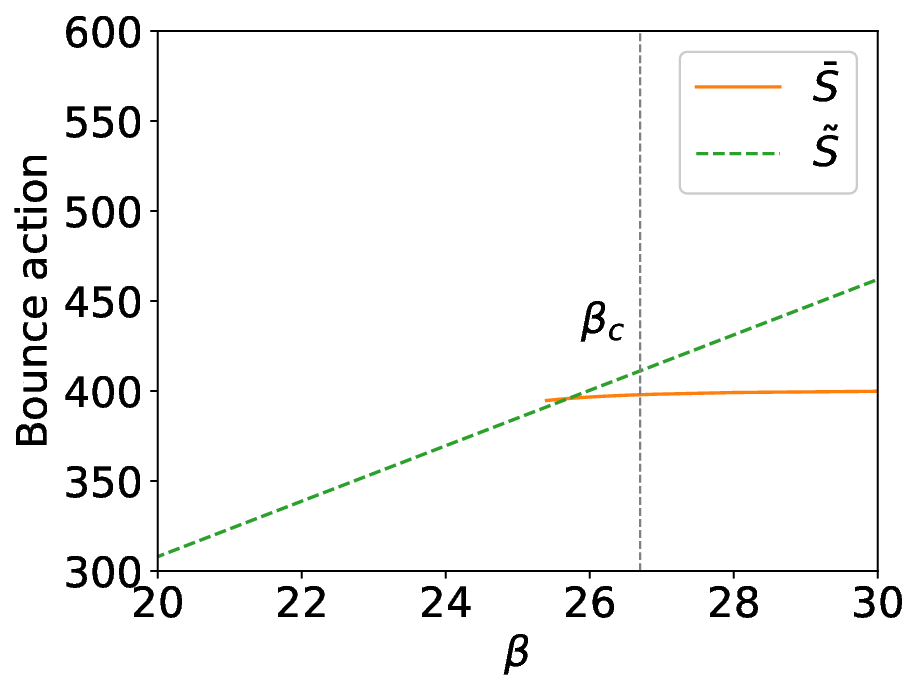}
    \end{minipage}
    \begin{minipage}{\linewidth}
        \centering
        \includegraphics[width=8cm]{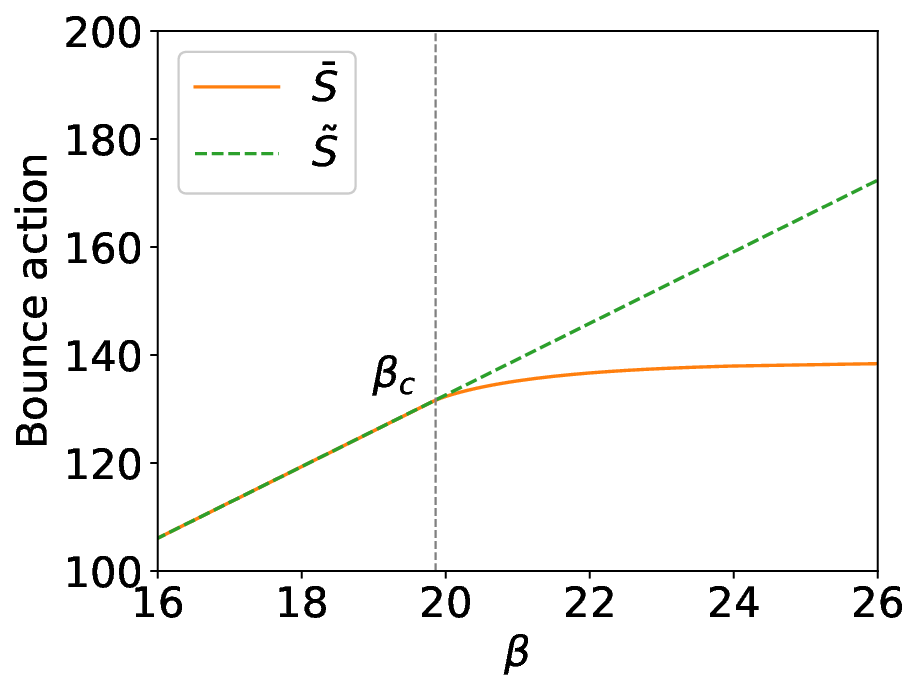}
    \end{minipage}
    \begin{minipage}{\linewidth}
        \centering
        \includegraphics[width=8cm]{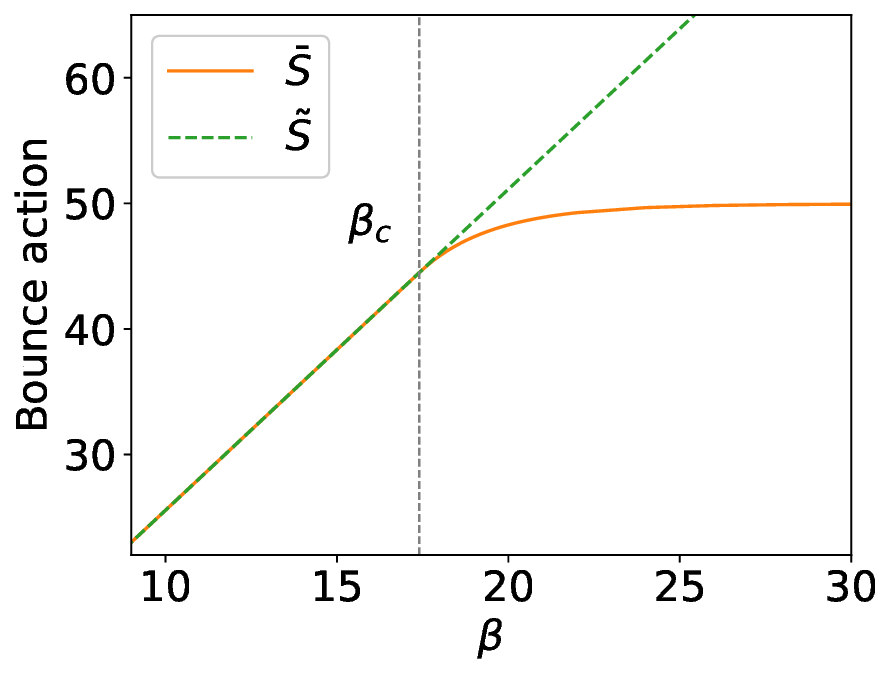}
    \end{minipage}
    \caption{$\bar{S}$ and $\tilde{S}$ as functions of $\beta$ for Model 4 with $\epsilon=0.15$ (top), Model 5 with $\epsilon=0.2$ (middle) and Model 6 with $\epsilon=0.25$ (bottom). The vertical dashed line shows $\beta_{\rm c}$.}
    \label{fig:Sethick}
\end{figure}

The bounce configurations for several temperatures are shown in Fig.\ \ref{fig:heatmapthick}. We can see that, as $\epsilon$ becomes larger, the thickness of the wall becomes comparable to the bounce size. When the temperature is so low that $\beta$ is much larger than the bounce size, the bounce shape is almost spherical in $D$-dimensional Euclidean space. As the temperature becomes higher, the bounce shape gradually changes. In particular, for Models 5 and 6, the $\tau$-dependent bounce continuously deforms into the $\tau$-independent bounce with the increase of the temperature; this is a big contrast to the thin-wall case. As in the case of the thin-wall, there only exists the $\tau$-independent bounce as the temperature becomes high enough. In the thick-wall case, the transition from the (almost) $D$-dimensional spherical bounce to the $\tau$-independent one occurs through the shape of the "wiggly cylinder" as discussed in Refs.\ \cite{Linde:1980tt, Linde:1981zj}. On the contrary, for the case of Model 4, the bounce shape looks similar to the case of thin-wall bounce; however, as we mentioned, the bounce action cannot be well approximated by the thin-wall approximation given in Eq.\ \eqref{Sbar(approx)}. Thus, we expect that Model 4 is corresponding to a case between the thin- and thick-wall limits.

In Fig.\ \ref{fig:Sethick}, we plot $\bar{S}$ and $\tilde{S}$ as functions of $\beta$.  As expected,  at low temperature, $\bar{S}$ is insensitive to $\beta$. On the contrary, the high-temperature behavior differs. In particular, for Models 5 and 6 which, we expect, well represent the thick-wall limit, $\bar{S}$ smoothly approaches to $\tilde{S}$ with the increase of the temperature; it is anticipated from the temperature dependence of the bounce configuration shown in Fig.\ \ref{fig:heatmapthick}.  Such a behavior of the bounce action in the thick-wall case was also indicated in Ref.\ \cite{Ferrera:1995gs}. 

In order to understand the properties of $\bar{S}$ and $\tilde{S}$, we also study $\beta_{\rm c}$, the critical inverse temperature below which the $\tau$-independent bounce is unstable against fluctuations. Using the method given in the previous subsection, we determine $\beta_{\rm c}$ for Models 4, 5 and 6; the result is shown in Fig.\ \ref{fig:Sethick} (vertical dashed line) as well as in Table \ref{table:thickwall}. Notably, in the thick-wall limit (i.e., Models 5 and 6), the lines of $\bar{S}$ and $\tilde{S}$ meet almost at $\beta_{\rm c}$ (see Fig.\ \ref{fig:Sethick}). This fact is a strong support of the expectation that, in the thick-wall limit, the $\tau$-dependent bounce continuously transforms into $\tau$-independent one with the increase of the temperature and that there is only one type of bounce contributing to the transition in a thermal bath.  The behavior in Model 4 is different because it is not in the thick-wall limit. In Model 4, the $\tau$-dependent bounce suddenly disappears at $\beta=\beta_*$; the $\tau$-dependent and $\tau$-independent bounces coexist at $\beta_*<\beta<\beta_{\rm c}$.

\subsection{Case of $D=2$}

\begin{table}[t]
  \begin{center}
    \begin{tabular}{l|rr}
      \hline\hline
      & Model 7 & Model 8 
      \\
      \hline
      $v$ & 1 & 1 
      \\
      $\lambda$ & $2.50\times 10^{-1}$ & $2.50\times 10^{-1}$
      \\
      $\epsilon$ & $2.50\times 10^{-2}$ & $2.50\times 10^{-1}$
      \\
      $\bar{S}$: Numerical & $6.93$& $1.33\times 10^{-1}$
      \\
      $\bar{S}$: {\tt CosmoTransitions} & $6.93$& $1.33\times 10^{-1}$
      \\
      $\bar{S}$: Eq.\ \eqref{Sbar(approx)} & $6.98$& $6.98\times 10^{-1}$
      \\
      $\beta_c$: Numerical& $2.42\times 10^{1}$& $1.78\times 10^{1}$
      \\
      \hline\hline
    \end{tabular}
    \caption{Model parameters used in our numerical analysis as well as the bounce parameters for the case of $D=2$. We take the unit such that $\lambda=0.25$. $\bar{S}$ is the bounce action in the low-temperature limit.}
    \label{table:results2d}
  \end{center}
\end{table}

Before closing this section, we comment on the case of $D=2$, which corresponds to the breaking of a string-like object (see, for example, Refs.\ \cite{Ivlev:1987zz, Preskill:1992ck}).  In such a case, the behavior of the bounce is expected to be altered because the friction term in Eq.\ \eqref{BounceEom} does not exist.  In Ref.\ \cite{Garriga:1994ut}, it was discussed that, when $D=2$, the transition from the $\tau$-dependent bounce to the $\tau$-independent one is continuous even in the thin-wall limit using semi-analytic argument.  The purpose of this subsection is to apply our numerical method the case of $D=2$ and to discuss the critical temperature $\beta_{\rm c}$ for the instability of the $\tau$-independent bounce. 

Here, we consider two models, corresponding to the thin-wall case (Model 7) and thick-wall case (Model 8).  Model parameters, as well as some of the bounce parameters, are summarized in Table \ref{table:results2d}. 

\begin{figure}[t]
    \centering
    \includegraphics[width=1\textwidth]{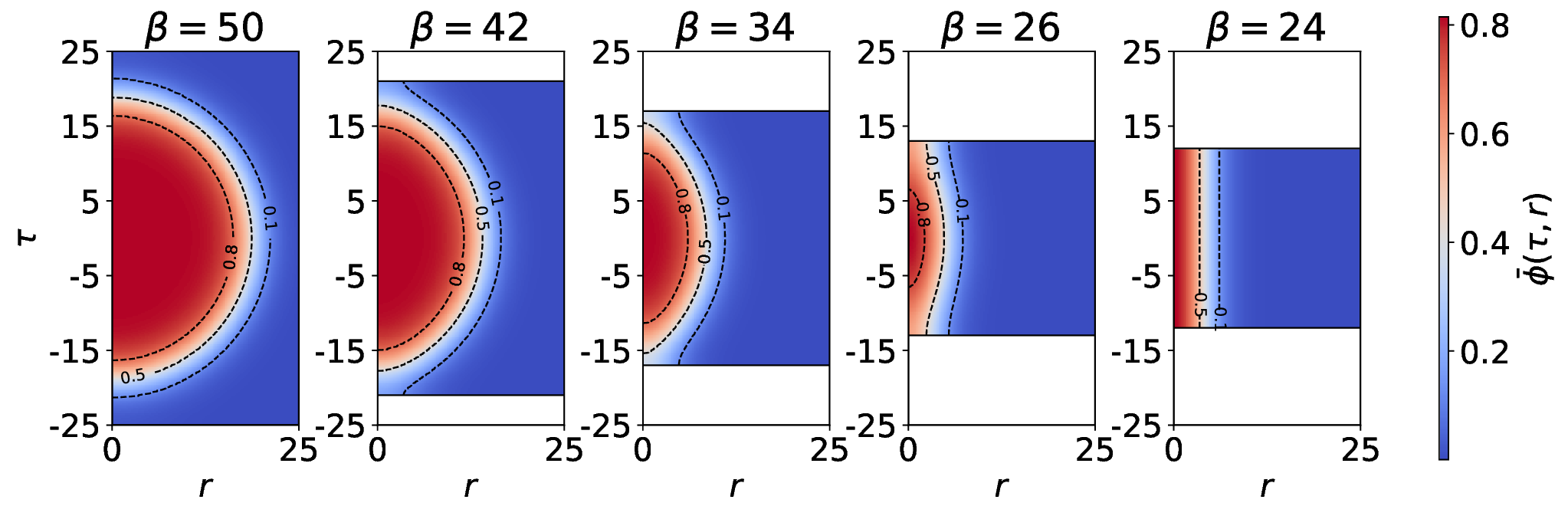}
    \includegraphics[width=1\textwidth]{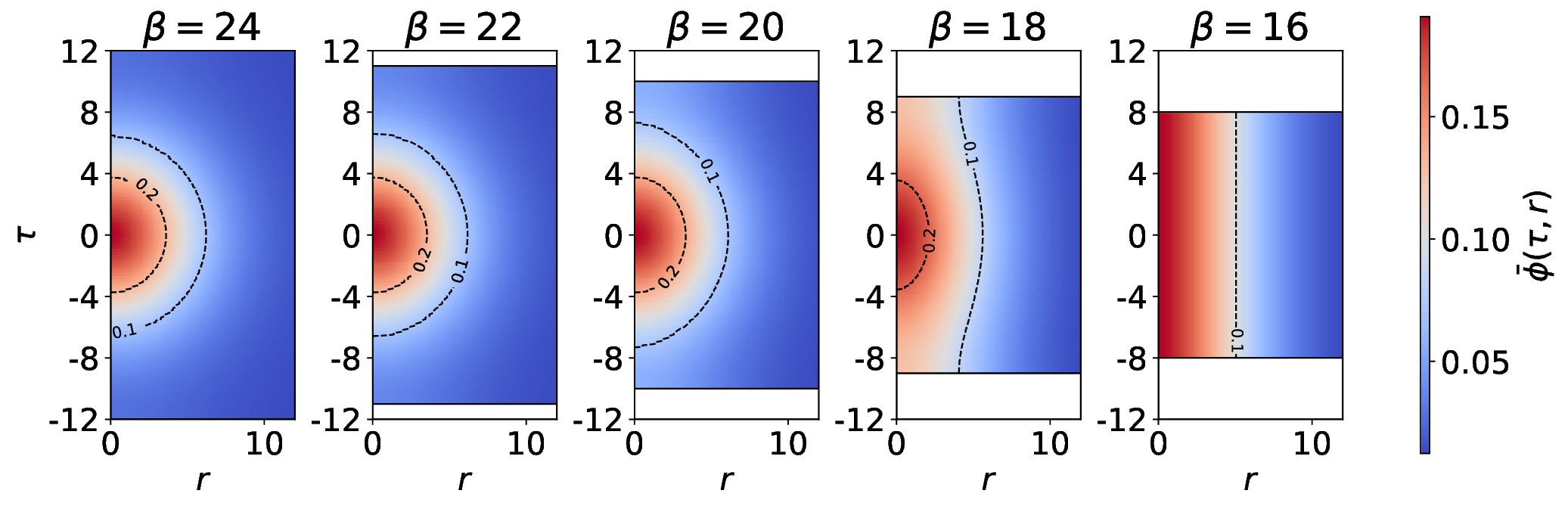}
    \caption{Bounce configuration for the $D=2$ cases: Model 7 with $\epsilon=0.025$ (top) and Model 8 with $\epsilon=0.25$ (bottom). The most left figures show $\tilde{\phi}$.}
    \label{fig:heat2d}
\end{figure}

Fig.\ \ref{fig:heat2d} shows the configuration of the bounce.  In the thin-wall limit, we can find a remarkable difference from the $D=4$ case. In the thin-wall case, even when $D=2$, the bounce shape is spherical at the low-temperature. However, with the increase of the temperature, the shape becomes non-spherical and it gradually approaches to that of the $\tau$-independent bounce.  The behavior is qualitatively the same for the thick-wall case.

\begin{figure}[t]
    \centering
    \begin{minipage}{\linewidth}
        \centering
        \includegraphics[width=7.5cm]{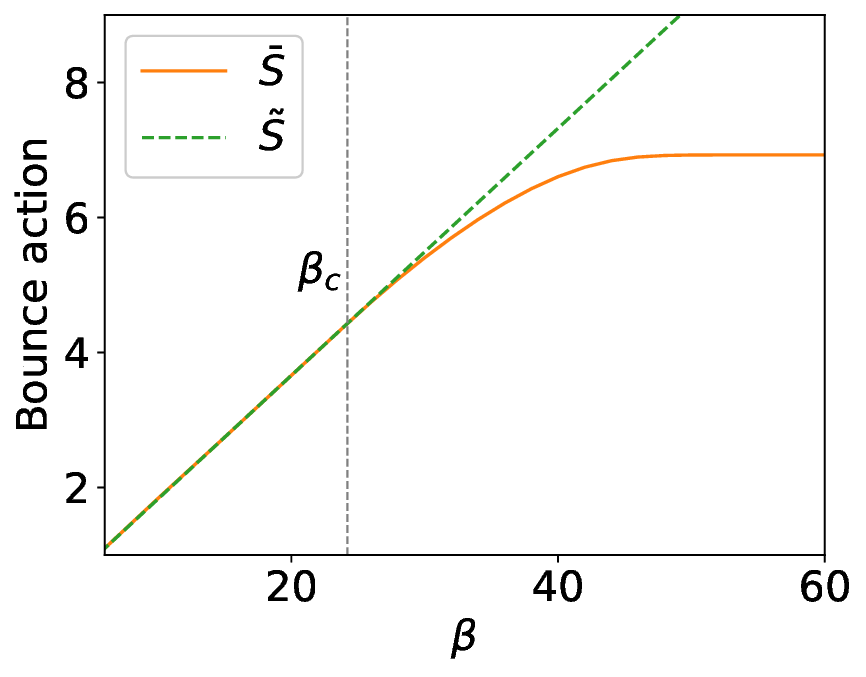}
        \includegraphics[width=8cm]{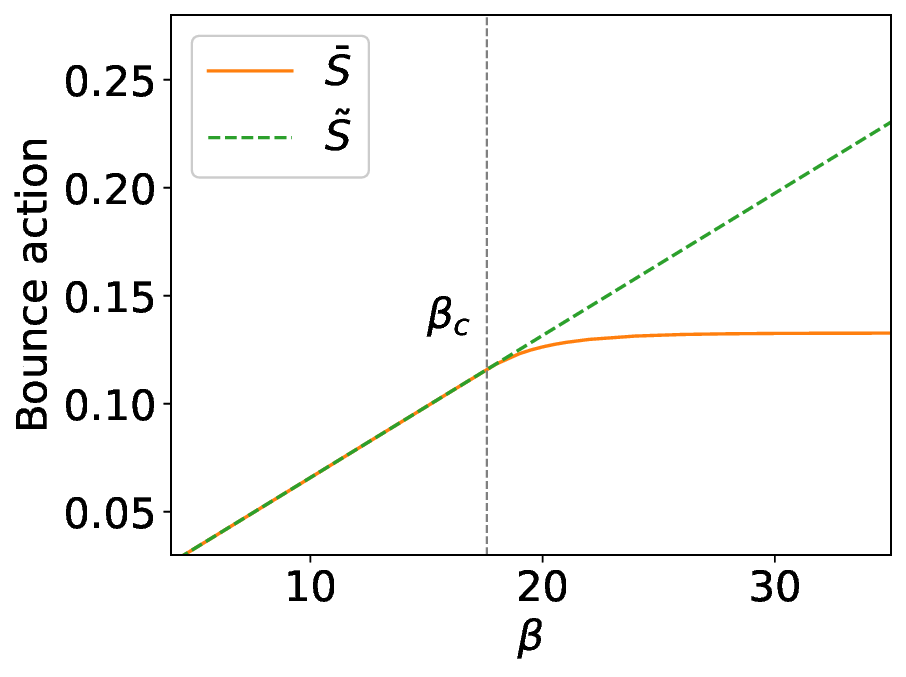}
    \end{minipage}
    \caption{$\bar{S}$ and $\tilde{S}$ as functions of $\beta$ for Model 7 with $\epsilon=0.025$ (left) and Model 8 with  $\epsilon=0.25$ (right). }
    \label{fig:Se2d}
\end{figure}

In Fig.\ \ref{fig:Se2d}, we show the $\beta$-dependence of the bounce action. As expected, in both thin- and thick-wall cases, the line of $\bar{S}$ merges to that of $\tilde{S}$. In addition, the critical inverse temperature $\beta_{\rm c}$ is indicated on the figure; we can find that the lines of $\bar{S}$ and $\tilde{S}$ meet (almost) at the inverse temperature of $\beta_{\rm c}$.  It indicates that the transition between the $\tau$-dependent and $\tau$-independent bounces is smooth at $\beta=\beta_{\rm c}$.

\section{Summary}
\label{sec:summary}
\setcounter{equation}{0}
\setcounter{figure}{0}

We have studied metastability of quantum fields in a thermal bath. Applying Affleck's analysis \cite{Affleck:1980ac} to the case of QFT, we investigated the transition from a metastable state to a state near the true vacuum. Our focus was on the detailed characterization of the bounce solution, which governs the transition rate. For the transition in a thermal bath, we should consider both $\tau$-dependent and $\tau$-independent bounces (denoted as $\bar{\phi}$ and $\tilde{\phi}$, respectively). We have studied both of them, paying particular attention to the $\beta$-dependence of bounce actions $\bar{S}$ and $\tilde{S}$.

The bounce for the transition in a thermal bath is defined on the Euclidean space-time with an introduced Euclidean time $\tau$, exhibiting periodicity in the $\tau$ direction with period $\beta$.  We have proposed a method to calculate the bounce with Fourier transforming the $\tau$-dependence of the bounce; our method can be implemented into numerical calculations.  

To validate our approach, we applied it across several model scenarios, numerically computing the bounce solutions and their actions. Our method accurately computed the bounces, confirmed by comparing numerical results with semi-analytic approaches.

For the transition in a thermal bath, we should consider two types of bounces, $\tau$-dependent bounce $\bar{\phi}$ and $\tau$-independent bounce $\tilde{\phi}$. We have studied the properties of $\bar{\phi}$ and $\tilde{\phi}$ in detail. The bounce having a non-trivial $\tau$-dependence does not exist at high enough temperature $\beta^{-1}>\beta_*^{-1}$.  In the case of the thin-wall case, $\bar{S}$ is larger than $\tilde{S}$ at $\beta\sim\beta_*$ and the transition from the $\tau$-dependent bounce to the $\tau$-independent one is discontinuous.  On the contrary, for the case of thick-wall, $\bar{S}$ smoothly approaches to $\tilde{S}$ at $\beta\sim\beta_*$.

An essential consideration is the stability of the $\tau$-independent bounce against fluctuations. At low temperature, the $\tau$-independent bounce is destabilized against fluctuation and is irrelevant. We have developed a method to determine the critical inverse temperature $\beta_{\rm c}$ below which such a destabilization occurs. We have developed a numerical procedure to calculate $\beta_{\rm c}$ to examine the stability of the $\tau$-independent bounce. For the thin-wall case, $\beta_*$ is significantly smaller than $\beta_{\rm c}$ and we expect that the $\tau$-dependent and $\tau$-independent bounces coexist for $\beta_*<\beta<\beta_{\rm c}$.  On the contrary, in the thick-wall case, our numerical calculation suggests $\beta_*=\beta_{\rm c}$.

\subsection*{Acknowledgments}
The work of Z.F. is supported by Forefront Physics and Mathematics Program to Drive Transformation (FoPM), a World-leading Innovative Graduate Study (WINGS) Program, the University of Tokyo.
The work of T.M. is supported by JSPS KAKENHI Grant No.\ 23K22486.

\appendix
\renewcommand{\thefigure}{\thesection.\arabic{figure}}

\section{Metastability in Quantum Mechanics}
\label{app:qm}
\setcounter{equation}{0}
\setcounter{figure}{0}

In this Appendix, we summarize the results of Ref.\ \cite{Affleck:1980ac} with extending the discussion to the case of the quantum-mechanical system with $N>1$ degrees of freedom.

\subsection{Quantum mechanical treatment of the transition rate}

In this subsection, we consider a quantum system with $N$ degrees of freedom. The canonical variable is denoted as $q_i$ ($i=1, \cdots, N$), which defines $N$-dimensional configuration space. The Lagrangian has the following form:
\begin{align}
  L = \frac{1}{2}  \dot{\bf q}^2 - U({\bf q})
  = \frac{1}{2} \sum_{i=1}^N \dot{q}_i^2 - U,
\end{align}
where ${\bf q}=(q_1, \cdots, q_N)$ is the set of canonical variables. The potential $U$ has two minima, corresponding to false and true vacua.\footnote
{The potential may have a run-away direction instead of the true vacuum.  The following arguments hold for the case with a run-away direction, so we do not distinguish two cases.}

\begin{figure}[t]
  \centering
  \includegraphics[width=0.5\linewidth]{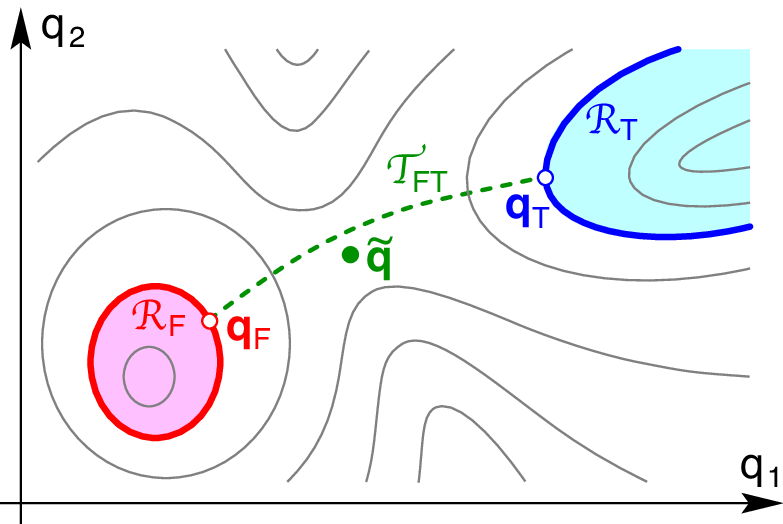}
  \caption{A schematic picture of the configuration space for the case of $N=2$.  The gray lines show the contours of constant $U$.  The pink- and skyblue-shaded regions are $\mathcal{R}_{\rm F}$ and $\mathcal{R}_{\rm T}$, respectively, while the thick red and blue lines are $\partial\mathcal{R}_{\rm F}$ and $\partial\mathcal{R}_{\rm T}$, respectively. The green dashed line indicates $\mathcal{T}_{\rm FT}$. The green dot is the saddle point $\widetilde{\bf q}$ of the potential $U$.}
  \label{fig:confspace}
\end{figure}

In the following, we consider the transition of the metastable state. For this purpose, we define two regions in $N$-dimensional configuration space, $\mathcal{R}_{\rm F}$ and $\mathcal{R}_{\rm T}$; $\mathcal{R}_{\rm F}$ is the region with $U<E$ containing the false vacuum, while $\mathcal{R}_{\rm T}$ is that containing the true vacuum.  Then, classically the particle with energy $E$ can exist in either $\mathcal{R}_{\rm F}$ or $\mathcal{R}_{\rm T}$.  Then, we consider the transition from $\mathcal{R}_{\rm F}$ to $\mathcal{R}_{\rm T}$ with energy $E$.  (A schematic picture of the configuration space for the case of $N=2$ is shown in Fig.\ \ref{fig:confspace}.)

The transition process can be studied with solving the Schr\"odinger equation:
\begin{align}
  E \Psi ({\bf q}) = 
  - \frac{\hbar^2}{2} \sum_{i=1}^N
  \frac{\partial^2}{\partial q_i^2} \Psi ({\bf q}) 
  + U ({\bf q}) \Psi ({\bf q}).
  \label{SchrodingerEq}
\end{align}
In this subsection, we keep $\hbar$ to explain the semiclassical approximation. We define the function $W$ with which the wave function is given as follows:
\begin{align}
  \Psi ({\bf q}) = e^{- W({\bf q}) / \hbar}.
\end{align}
Substituting the above expression to Eq.\ \eqref{SchrodingerEq} and keeping only the leading-order term with respect to $\hbar$ adopting the semiclassical approximation, the function $W$ is found to obey
\begin{align}
  - E = \frac{1}{2} \sum_{i=1}^N \left( \frac{\partial W}{\partial q_i} \right)^2 - U.
  \label{HamiltonJacobi}
\end{align}
In order to discuss the transition from the state in $\mathcal{R}_{\rm F}$ to that in $\mathcal{R}_{\rm T}$, we study the behavior of the wave function in the region of $U>E$ (i.e., the region of the potential barrier).  

\subsection{Hamilton-Jacobi equation}

Eq.\ \eqref{HamiltonJacobi} is nothing but Hamilton-Jacobi (HJ) equation for a classical particle with energy $-E$ propagating in $N$-dimensional space with potential $-U$. The EoM of such a particle is given by
\begin{align}
  \ddot{q}_i = \frac{\partial U}{\partial q_i},
  \label{EoM}
\end{align}
and the conservation law requires
\begin{align}
  E = U - \frac{1}{2} \sum_{i=1}^N \dot{q}_i^2,
  \label{EnergyConservation}
\end{align}
where the ``dot'' denotes the derivative with respect to time.  Obviously, in this case, the particle can propagate the region with $E<U$.

A solution of the HJ equation \eqref{HamiltonJacobi} describes a set of solutions of classical EoM, given by Eq.\ \eqref{EoM}, satisfying the constraint given in Eq.\ \eqref{EnergyConservation} \cite{Landau:1982Mechanics}. Each solution of the classical EoM gives one trajectory of propagation, and the function $W$ gives various solutions with different trajectories. For a given $W$, the canonical momentum of the particle at ${\bf q}$ is given by the gradient of $W$:
\begin{align}
  p_i = \dot{q}_i = \frac{\partial W}{\partial q_i}.
\end{align}
Thus, the constant-$W$ hypersurface is perpendicular to the classical trajectories (and hence it is perpendicular to ${\bf p}$). Once a set of classical solutions in association with $W$ are known, the function $W$ can be reproduced as
\begin{align}
  W ({\bf q}) = \int^{\bf q} d{\bf q}\, {\bf p} ({\bf q}).
  \label{intdqp}
\end{align}
Here, the line integral is along the classical trajectory $\mathcal{T}(\bf q)$ which goes through the point ${\bf q}$.

Hereafter, we consider the case that there exists a classical solution of Eq.\ \eqref{EoM} connecting a point ${\bf q}_{\rm F}\in\partial\mathcal{R}_{\rm F}$ and another point ${\bf q}_{\rm T}\in\partial\mathcal{R}_{\rm T}$, where $\partial\mathcal{R}$ denotes the boundary of the region $\mathcal{R}$. Because of Eq.\ \eqref{intdqp}, the function $W$ for such a solution should satisfy
\begin{align}
  W ({\bf q}_{\rm T}) - W ({\bf q}_{\rm F}) = 
  \int_{\mathcal{T}_{\rm FT}} d{\bf q}\, {\bf p} =
  \int_{\mathcal{T}_{\rm FT}} dq \sqrt{2(U - E)},
\end{align}
where $dq={\bf e}_{\bf p}d{\bf q}$ (with ${\bf e}_{\bf p}$ being the unit vector pointing to the direction of ${\bf p}$) denotes the line integral along the classical path from ${\bf q}_{\rm F}$ to ${\bf q}_{\rm T}$.

Once the particle starts its motion from ${\bf q}={\bf q}_{\rm F}$ (with $\dot{\bf q}=0$), it oscillates as ${\bf q}_{\rm F}\rightarrow{\bf q}_{\rm T}\rightarrow{\bf q}_{\rm F}$; such a solution of Eq.\ \eqref{EoM} is called as ``bounce,'' and is denoted as $\bar{\bf q}(t)$ hereafter.  The oscillation period of the bounce is given by
\begin{align}
  \bar{\tau} \equiv 2 \int_{\mathcal{T}_{\rm FT}} dq \frac{1}{\sqrt{2(U - E)}}.
\end{align}
Using the translational invariance to the time direction, we take $\bar{\bf q}(0)=\bar{\bf q}(\bar{\tau})={\bf q}_{\rm F}$ and $\bar{\bf q}(\bar{\tau}/2)={\bf q}_{\rm T}$. The potential $-U$ is unbounded from below and the bounce solution $\bar{\bf q}$ is unstable against fluctuations. 

\subsection{Transition rate: quantum-mechanical estimation}

The arguments so far can be utilized in the study of the transition process. Assuming that the wave function in the potential barrier is governed by the function $W$ describing the classical trajectory $\mathcal{T}_{\rm FT}$, we expect that the wave function has the following property:
\begin{align}
  \Psi ({\bf q}_{\rm T}) \sim
  e^{-w(E)} \Psi ({\bf q}_{\rm F}),
\end{align}
with
\begin{align}
  w(E) \equiv
  \int_{\mathcal{T}_{\rm FT}} dq \sqrt{2(U - E)}.
\end{align}
(Here and hereafter, we use the $\hbar=1$ unit for notational simplicity.) We neglect subleading-order effects in semiclassical approximation; the above expression is enough to estimate the quantity $B$.

At ${\bf q}\sim {\bf q}_{\rm F/T}$, the function $W$ depends only on $q_\perp$, where $q_\perp$ is the coordinate variable (given by a linear combination of $q_i$'s) for the direction perpendicular to $\partial\mathcal{R}_{\rm F/T}$. In order to obtain the wave function in the region of $E>U$ (i.e., in $\mathcal{R}_{\rm F}$ and $\mathcal{R}_{\rm T}$) at around ${\bf q}\sim {\bf q}_{\rm F/T}$, we may neglect the dependence on variables other than $q_\perp$ and use the matching condition for the oscillating and damping regimes in semiclassical approximation (see, for example, \cite{Landau:1991wop}).  Then, for ${\bf q}\in \mathcal{R}_{\rm F}$ and ${\bf q}\in \mathcal{R}_{\rm T}$, we obtain 
\begin{align}
  \Psi ({\bf q}\sim {\bf q}_{\rm F}) \sim 
  \Psi ({\bf q}_{\rm F}) 
  \cos \left[ \int_{q_\perp}^{q_{\rm F}} dq_\perp \sqrt{2(E-U)} - \frac{\pi}{4} \right] ~~~ (\mbox{for}\, {\bf q}\in \mathcal{R}_{\rm F}),
\end{align}
and 
\begin{align}
  \Psi ({\bf q}\sim {\bf q}_{\rm T}) \sim 
  \Psi ({\bf q}_{\rm T}) 
  \exp 
  \left[ i \left( \int_{q_{\rm T}}^{q_\perp} dq_\perp \sqrt{2(E-U)} - \frac{\pi}{4} \right) \right] ~~~ (\mbox{for}\, {\bf q}\in \mathcal{R}_{\rm T}),
\end{align}
where $q_\perp=q_{\rm F/T}$ at $\partial\mathcal{R}_{\rm F/T}$, and $q_\perp<q_{\rm F}$ ($q_\perp>q_{\rm T}$) in ${\bf q}\in\mathcal{R}_{\rm F}$ (${\bf q}\in\mathcal{R}_{\rm T}$). As one can see, $\Psi ({\bf q}\sim {\bf q}_{\rm F})$ consists of waves incoming to and reflecting from the potential barrier, while $\Psi ({\bf q}\sim {\bf q}_{\rm T})$ describes the wave after the tunneling.  Postulating that the tunneling is governed by the behaviors of the wave function at ${\bf q}={\bf q}_{\rm F}$ and ${\bf q}_{\rm T}$, we estimate the tunneling rate from the ratio of the incoming and tunneled fluxes as
\begin{align}
  \gamma (E) \sim \frac{|\Psi ({\bf q}_{\rm F})|^2}{|\Psi ({\bf q}_{\rm T})|^2}
  \sim e^{-2w(E)}.
\end{align}

Now, we consider the transition in a thermal bath with the inverse temperature $\beta$. The transition rate is evaluated by thermally averaging $\gamma (E)$ as \cite{Affleck:1980ac}
\begin{align}
  \Gamma = \int dE D(E) \gamma (E) e^{-\beta E}
  \sim \int dE D(E) e^{-2w (E) - \beta E},
  \label{gammabar}
\end{align}
where $D(E)$ is the state density. We evaluate $\Gamma$ by the steepest descent method, assuming that the energy dependence of $D(E)$ is mild enough. The integration is dominated by the energy satisfying $2\frac{dw}{dE}+\beta=0$, which gives
\begin{align}
  \bar{\tau} (E) = \beta,
  \label{tau=beta}
\end{align}
thus the classical solution having the oscillation period equal to $\beta$ is important. Existence of such a solution depends on $E$.
\begin{itemize}
\item[(a)] If there exists a solution satisfying Eq.\ \eqref{tau=beta}, which we denote $\bar{\bf q}_\beta$, the decay rate is estimated as $\Gamma \sim e^{-2w-\bar{\tau} E}|_{\bar{\tau}=\beta}$.  Using the fact that $2w+\bar{\tau} E$ is the Euclidean action of of the bounce, we can find
\begin{align}
  \Gamma \sim e^{-S_\beta},
  \label{gammabar_lowT}
\end{align}
where $S_\beta$ is the action of the bounce having the oscillation period equal to $\beta$:
\begin{align}
  S_\beta \equiv \int_0^\beta dt 
  \left[ \frac{1}{2} \dot{\bar{\bf q}}_\beta^2 + U (\bar{\bf q}_\beta) \right].
\end{align}
\item[(b)] As $E$ gets larger, ${\bf q}_{\rm F}$ and ${\bf q}_{\rm T}$ both become close to the saddle point of the potential; we denote the saddle point as $\widetilde{\bf q}$, which satisfies 
\begin{align}
  \left. \frac{\partial U}{\partial q_i} \right|_{{\bf q}=\widetilde{\bf q}} = 0.
  \label{hatq}
\end{align}
We approximate the potential around the saddle point as
\begin{align}
  U = \widetilde{U}
  + \frac{1}{2} \sum_{ij} \widetilde{U}_{ij} (q_i - \widetilde{q}_i) (q_j - \widetilde{q}_j) + \cdots.
\end{align}
The matrix $\widetilde{U}_{ij}$ has one negative eigenvalue which we denote $-\mu_{}^2$. When $E\rightarrow \widetilde{U}$ (with $E<\widetilde{U}$), the bounce is approximately given by a harmonic oscillation to the direction of the eigenstate of $\widetilde{U}_{ij}$ with the negative eigenvalue, so the bounce period becomes close to $2\pi/\mu$. This implies that $\bar{\tau}$ is bounded from above. When $\beta<2\pi/\mu$, the thermal average of the transition rate cannot be evaluated by the steepest descent method and the integral introduced in Eq.\ \eqref{gammabar} is dominated by $E\sim \widetilde{U}$. Then, we expect
\begin{align}
  \Gamma \sim e^{-\beta \widetilde{U}}.
  \label{gammabar_highT}
\end{align}
\end{itemize}
Notice that the case (b) is often called as ``thermal hopping.'' 

As pointed out in Ref.\ \cite{Affleck:1980ac}, the exponential factors arising in Eqs.\ \eqref{gammabar_lowT} and \eqref{gammabar_highT} are equal to those showing up in evaluating ${\rm Im}F$ by the path-integral formulation. Based on such an observation, the transition rate in a thermal bath is considered to be proportional to ${\rm Im}F$.\footnote
{In Ref.\ \cite{Affleck:1980ac}, the effect of the prefactor in front of the exponential factor was also studied; in particular, for the case of $N=1$, it has been also shown that the transition rate is given by $(\mu\beta/\pi){\rm Im}F$ with calculating the prefactor.}

\section{Numerical Algorithm for Calculating the Bounce}
\label{app:algorism}
\setcounter{equation}{0}
\setcounter{figure}{0}

In this Appendix, we introduce our numerical algorithm for solving Eq.\ \eqref{EqForNumericalCalc}. Here, for simplicity, we concentrate on the case with a single field; generalizations to the cases with a multiple of scalar fields are straightforward. In the following, the following scalar potential is considered:
\begin{align}
  V = g^{(1)} \phi
  + \frac{1}{2} g^{(2)} \phi^2
  + \frac{1}{3} g^{(3)} \phi^3
  + \frac{1}{4} g^{(4)} \phi^4.
\end{align}

We start with the fact that the bounce solution $\bar{\phi}$ satisfies the following differential equation:
\begin{align}
  - \left( \partial_\tau^2 + \partial_r^2 + \frac{D-2}{r} \partial_r \right) \bar{\phi}
  + \left. \frac{\partial V}{\partial \phi} \right|_{\phi=\bar{\phi}} = 0.
  \label{EqBounce}
\end{align}
We discretize the above equation and solve it with the Newton's method.

Because we are interested in functions having the same periodicity and boundary conditions as the bounce, we expand the function $\phi$ as
\begin{align}
  \phi (\tau, r) = &\,
  \sum_{n=0}^{N_\tau} \varphi_{n} (r)
  \cos \left( \frac{2n\pi}{\beta} \tau \right).
  \label{phi_numerical}
\end{align}
As we will confirm by the numerical calculation, the bounce does not have a structure smaller than the wall width, so we expect that Eq.\ \eqref{phi_numerical} well describes the bounce solution if $N_{\rm \tau}$ is large enough. The boundary conditions are given by
\begin{align}
  \partial_r \varphi_n (r=0)=0,~~~\varphi_n(r\to \infty)=v_{\rm F}\delta_{n,0},
  \label{EqForFourierBounD}
\end{align}
where $v_{\rm F}$ is the amplitude of $\phi$ at the false vacuum. We also discretize the $r$-variable as
\begin{align}
    r_I & \equiv r_0+ I \delta r~~~
    (I=0, 1, \cdots, N_r),
    \label{r_I}
\end{align}
where $r_0$ and $\delta r$ are parameters which are much smaller than the bounce size, and define
\begin{align}
  \varphi_{n,I} \equiv \varphi_n(r_I).
\end{align}
The parameters $\delta r$ and $N_r$ are chosen such that $r_{\rm max}$ is much larger than the bounce size, where
\begin{align}
  r_{\rm max} \equiv r_0 + N_r \delta r.
  \label{r_max}
\end{align}
Our purpose is to determine $\varphi_{n,I}$'s which describe the bounce solution.

In order to use the Newton's method, we introduce functions $f_{n,I}$'s which depend on $\varphi_{n,I}$'s and vanish when $\varphi_{n,I}$'s give the bounce.  For $1\leq I\leq N_r-1$, $f_{n,I}$ is given by discretizing Eq.\ \eqref{EqBounce}:
\begin{align}
  f_{n,I} \equiv &\, 
  - \left(
  \frac{\varphi_{n,I+1} - 2\varphi_{n,I} + \varphi_{n,I-1}}{\delta r^2}
  + \frac{D-2}{r_I} \frac{\varphi_{n,I+1} - \varphi_{n,I-1}}{2 \delta r}
  \right)
  + \left(\frac{2n \pi}{\beta}\right)^2 \varphi_{n,I}
  \nonumber \\ &\, 
  + g^{(1)}\delta_{n,0}
  + g^{(2)} \varphi_{n,I}
  + g^{(3)} \sum_{\ell_1, \ell_2} C_{n, \ell_1, \ell_2, 0} \varphi_{\ell_1, I} \varphi_{\ell_2, I} 
  + g^{(4)} \sum_{\ell_1, \ell_2, \ell_3} C_{n, \ell_1, \ell_2, \ell_3} \varphi_{\ell_1, I} \varphi_{\ell_2, I} \varphi_{\ell_3, I},
\end{align}
where the second-order accurate central-difference scheme is used to express the derivative with respect to $r$. In addition, $f_{n,0}$ and $f_{n,N_r}$ are introduced so that the function $\phi$ satisfies the boundary conditions \eqref{EqForFourierBounD}. The boundary condition at $r=0$ is imposed with introducing 
\begin{align}
  f_{n,0} \equiv 3\varphi_{n,0}-4\varphi_{n,1}+\varphi_{n,2},
  \label{f_n0}
\end{align}
which is due to the relation $\partial_r\phi (r_0)\simeq \partial_r\phi (r_1)-\partial^2_r\phi (r_1)\Delta r \simeq 0$. The boundary condition at $r\rightarrow\infty$ is taken into account with
\begin{align}
  f_{n,N_r} \equiv \varphi_{n,N_r} - v_{\rm F} \delta_{n,0}.
\end{align}

Now, we introduce
\begin{align}
  \bm{F} ( \bm{\varphi} )
  \equiv \left(f_{0,0}, \cdots,f_{n,I}, \cdots,f_{N_\tau,N_r}\right),
\end{align}
where
\begin{align}
  \bm{\varphi} \equiv
  \left(\varphi_{0,0}, \cdots,\varphi_{n,I}, \cdots,\varphi_{N_\tau,N_r}\right).
\end{align}
The bounce solution can be obtained by solving
\begin{align}
  \bm{F}\left(\bm{\varphi}\right)=0.
  \label{F=0}
\end{align}

We solve Eq.\ \eqref{F=0} using Newton's method in multiple dimensions. Newton's method is an iterative technique used to find successively better approximations to the roots (or zeros) of a real-valued function. In the context of multiple dimensions, this method employs the Jacobian matrix to update the solution estimate at each step. By solving a linear system involving the Jacobian and the function values, Newton's method converges quadratically to the solution under suitable conditions. Our iteration equation is:
\begin{align}
  \bm{\varphi}^{(\nu+1)}= \bm{\varphi}^{(\nu)}
  - \mathcal{M}\left(\bm{\varphi}^{(\nu)}\right)^{-1}\bm{F}\left(\bm{\varphi}^{(\nu)}\right),
\end{align}
where $\nu$ is the iteration index and $\mathcal{M}$ is the Jacobian matrix defined as:
\begin{align}
  \mathcal{M}_{(n,I),(m,J)} \equiv
  \frac{\partial f_{n,I}}{\partial\varphi_{m,J}}.
\end{align}
Instead of inverting $\mathcal{M}$, we may use the Gaussian elimination method to find $\delta \bm{\varphi}^{(\nu)}$ that satisfies:
\begin{align}
  \mathcal{M}\left(\bm{\varphi}^{(\nu)}\right)\delta \bm{\varphi}^{(\nu)} = \bm{F}\left(\bm{\varphi}^{(\nu)}\right),
\end{align}
with which
\begin{align}
  \bm{\varphi}^{(\nu+1)} = \bm{\varphi}^{(\nu)} - \delta \bm{\varphi}^{(\nu)}.
\end{align}
The magnitude \(|\delta \bm{\varphi}_n|\) is defined as the error of our iteration. In our numerical analysis, we terminate our iteration when $|\delta \bm{\varphi}_n|<10^{-7}v$.

We note here that the procedure to calculate the bounce solution given above is also applicable to find the $\tau$-independent bounce. In particular, if we start with an initial guess such that $\varphi_{n\neq 0}^{(0)}=0$, then $\varphi_{n\neq 0}$'s always vanish during the iteration.

Our numerical calculation to find the bounce solution in a thermal bath proceeds as follows:
\begin{itemize}
\item[1.] For a given scalar potential, we first determine the zero-temperature bounce by the method explained above.
\item[2.] The zero-temperature bounce is used as an initial guess for the Newton's method for the case of $\beta=2r_{\rm max}\equiv\beta_0$ (which is much larger than the bounce size during our analysis). The amplitude for the first iteration $\bm{\varphi}^{(0)}$ is determined with Fourier-transforming and discretizing the zero-temperature bounce. Then, the bounce solution for $\beta=\beta_0$ is obtained by the Newton's method.
\item[3.] Using the bounce solution for $\beta=\beta_0$ as an initial guess, the bounce solution at a slightly higher temperature is obtained. 
\item[4.] The step of increasing the temperature is repeated; the bounce solution at a lower temperature is used as an initial guess to calculate the bounce at the temperature one step higher.
\end{itemize}
%
%
The numerical code used in our calculation is detailed in Appendix \ref{app:code}.
By following this procedure, we can obtain solutions across a range of temperature, from zero to high temperature.  In our numerical calculation, we take $N_\tau=60$ irrespective of $\beta$. The values of $r_{\rm max}$ and $N_r$ are varied depending on the Model.  For the $D=4$ case, we take $(r_{\rm max},N_r)=(95, 950)$, $(50, 1000)$, and $(30, 600)$, for Models 1, 2 and 3, respectively, while $r_{\rm max}=25$ and $N_r=500$ for thick-wall models. For the case of $D=2$, we slightly change the method of the numerical calculation, with considering the region $-\infty<r<\infty$ and imposing the boundary condition $\bar{\phi}(\tau,\pm\infty)=v_{\rm F}$.  The definition of the parameter $r_I$ given in Eq.\ \eqref{r_I} as well as the boundary condition \eqref{f_n0} are modified accordingly; in particular, $r_I=-r_{\rm max}+I\delta r$ with $\delta r=2r_{\rm max}/N_r$. We take $(r_{\rm max},N_r)=(35, 700)$ and $(25, 500)$ for Models 7 and 8, respectively.

\section{Numerical Code}
\label{app:code}
\setcounter{equation}{0}
\setcounter{figure}{0}

In this Appendix, we introduce the code used in our analysis, which numerically realize the procedure explained in Appendix \ref{app:algorism}. The following Python modules are used in our code:
\begin{itemize}
    \item \texttt{numpy} (imported as \texttt{np}),
    \item \texttt{scipy.interpolate} (imported as \texttt{interp1d}),
    \item \texttt{scipy.integrate} (imported as \texttt{simpson}).
\end{itemize}
The code is divided into three parts, each of which is detailed in the following subsections. 

\subsection{Zero-temperature bounce}

First, we find the zero-temperature bounce, which depends only on the radius in the $D$-dimensional Euclidean space and is $O(D)$ symmetric. The $O(D)$ symmetric bounce is used as the initial guess for the iteration to find the bounce in a thermal bath. The field equation is given by
\begin{align}
  - \left( \partial_r^2 + \frac{D-1}{r} \partial_r \right) \bar{\phi}
  + \left. \frac{\partial V}{\partial \phi} \right|_{\phi=\bar{\phi}} = 0,
  \label{EqBounceZeroT}
\end{align}
where $r$ in this subsection denotes the radius in the $D$-dimensional Euclidean space. (It is related to the spacial radius used in the main part of this paper, denoted as $r_{D-1}$ here, as $r=\sqrt{\tau^2+r_{D-1}^2}$.)  In addition, the boundary conditions are
\begin{align}
    \partial_r {\bar{\phi}}(r=0)=0, \ \bar{\phi}(r\to \infty)=0.
    \label{BoundaryConditionZeroT}
\end{align}
We discretize the $r$-variable as Eq.\ \eqref{r_I}, and define 
\begin{align}
  \varphi_{I} \equiv \phi(r_I).
\end{align}
The parameters $\delta r$ and $N_r$ are chosen so that $r_{\rm max}$ given in Eq.\ \eqref{r_max} is much larger than the bounce size.

To use Newton's method, we introduce functions $f_{I}$ that depend on $\varphi_{I}$ and become zero when $\varphi_{I}$ corresponds to the bounce. For $1 \leq I \leq N_r-1$, $f_{I}$ is defined by discretizing Eq.\ \eqref{EqBounceZeroT}:
\begin{align}
  f_{I} \equiv &\, 
  - \left(
  \frac{\varphi_{I+1} - 2\varphi_{I} + \varphi_{I-1}}{\delta r^2}
  + \frac{D-1}{r_I} \frac{\varphi_{I+1} - \varphi_{I-1}}{2 \delta r}
  \right)
  \nonumber \\ &\, 
   +\varphi_{I} \left(2 \lambda  v^2-6 \lambda  v^2 \epsilon \right)+ \varphi_{I}^2 (6 \lambda  v \epsilon -6 \lambda  v)+4 \lambda   \varphi_{I}^3.
   \label{f_I}
\end{align}
Additionally, $f_{0}$ and $f_{N_r}$ are introduced to ensure that $\bar{\phi}$ satisfies the boundary conditions \eqref{BoundaryConditionZeroT}. The boundary condition at $r=0$ is implemented by introducing 
\begin{align}
  f_{0} \equiv 3\varphi_{0}-4\varphi_{1}+\varphi_{2}.
  \label{f_0}
\end{align}
The boundary condition at $r\rightarrow\infty$ is handled with
\begin{align}
  f_{N_r} \equiv 0.
\end{align}

We define
\begin{align}
  \bm{F} ( \bm{\varphi}_{O(D)} )\equiv \left(f_{0}, \cdots,f_{N_r}\right),
\end{align}
where
\begin{align}
  \bm{\varphi}_{O(D)} \equiv \left(\varphi_{0}, \cdots,\varphi_{N_r}\right).
\end{align}
Then, the $O(D)$ symmetric bounce solution is obtained by solving
\begin{align}
  \bm{F}\left(\bm{\varphi}_{O(D)}\right)=0.
\end{align}
In the numerical code, the Newton method is adopted to solve the above equation, as explained in Appendix \ref{app:algorism}

We define $\delta \bm{\varphi}_{O(D)}^{(\nu)}$ as
\begin{align}
  \delta \bm{\varphi}_{O(D)}^{(\nu)}\equiv \bm{\varphi}_{O(D)}^{(\nu)} -\bm{\varphi}_{O(D)}^{(\nu+1)},
\end{align}
where $\bm{\varphi}_{O(D)}^{(\nu)}$ denotes the result after $\nu$-th iteration. The magnitude \(|\delta \bm{\varphi}_{O(D)}^{(\nu)}|\) represents the error of our iteration. In our numerical analysis, we stop the iteration when $|\delta \bm{\varphi}_{O(D)}^{(\nu)}|<10^{-7}v$. 

The initial guess $\bm{\varphi}_{O(D)}^{(0)}$ is crucial for convergence. For the thin-wall case ($\epsilon \ll 1$), the following function is suitable:
\begin{align}
  \bm{\varphi}_{O(D)}^{(0)}(r)\simeq \frac{1}{2} 
  \left[ 1 - \tanh \left\{ \frac{1}{2} m_\phi (r - \bar{r}) \right\} \right]
  v,
\end{align}
where $m_\phi$ and $\bar{r}$ are approximated as
\begin{align}
    m_\phi &\simeq \sqrt{2 \lambda} v,
\end{align}
and
\begin{align}
    \bar{r} & \simeq \frac{D-1}{3 \epsilon}m^{-1}_\phi.
    \label{rbar}
\end{align}
For the thick-wall case, a value of $\bar{r}$ close to expression (\ref{rbar}) often gives a convergent result. 

The following code is used to find the $O(D)$ symmetric solution $\bm{\varphi}_{O(D)}$. In using the code for a given potential, variables with ``...'' should be replaced with specific values.
\begin{lstlisting}[label={firstpart}]
# Definition of functions used in Jacobian M and the vector field F
def V(x): 
def dV(x): 
def ddV(x): 
def f(x,y,z,r): 
# Definition of parameters
lamb=...
epsi=...
v=...
rmax=...
Nr=...
D=...
varphiO4=[0]*(Nr+1)
r=[0]*(Nr+1)
rdelta=rmax/Nr
mphi=pow(2*lamb,1/2)*v
rbar=(D-1)/3/epsi/mphi*v
# Discretize r and set the initial guess for varphi
for i  in range (0,Nr+1):
    r[i]=pow(10,-6)+(i)*rdelta
    varphiO4[i]=v/2*np.tanh((-r[i]+rbar)/2/mphi)+v/2
# Itetarion
print("finding the zero-temperature bounce")
for m in range (10):
    F=[0]*(Nr+1)
    F[0]=3*varphiO4[0]-4*varphiO4[1]+varphiO4[2]
    F[Nr]=varphiO4[Nr]
    M=np.zeros((Nr+1, Nr+1))
    M[0,0]=3
    M[0,1]=-4
    M[0,2]=1
    M[Nr,Nr]=1
    for i in range (1,Nr):
        F[i]=f(varphiO4[i+1],varphiO4[i],varphiO4[i-1],r[i])
        M[i,i+1]=1+(D-1)/r[i]*rdelta/2
        M[i,i]=-2-pow(rdelta,2)*ddV(varphiO4[i])
        M[i,i-1]=1-(D-1)/r[i]*rdelta/2
    varphiO4=varphiO4-np.dot(np.linalg.inv(M),F)
    varphidelta=np.dot(np.linalg.inv(M),F)
    print("norm of error=",np.linalg.norm(varphidelta),"norm of F=",np.linalg.norm(F))
    if np.max(np.abs(varphidelta))<v*pow(10,-7):
        print("zero-temperature bounce has been founded")
        break
\end{lstlisting}

\subsection{Setting the initial guess}

After obtaining the zero-temperature bounce, we place it on the $\tau$ vs.\ $r$ plane. We first interpolate the discrete $\bm{\varphi}_{O(D)}$ using the cubic spline interpolation; the continuous function obtained from $\bm{\varphi}_{O(D)}$ is denoted as $\bar{\phi}_{O(D)}$. We descretize the $\tau$ direction to match the $O(D)$ bounce and the initial guess.  The number of the points in the $\tau$ direction, denoted as $N_\tau$, should be larger than $\beta/d$ (with $d$ being the wall width). $\tau$ is discretized as 
\begin{align}
    \tau_I \equiv -\frac{\beta}{2}+ I \delta t~~~
    (I=0, 1, \cdots, N_r),
    \label{tau_I}
\end{align}
where $\delta t \equiv \beta/N_r$. For discrete points $(r_I,\tau_J)$ on the $\tau$ vs.\ $r$ plane, the initial guess for the iteration (denoted as $\bar{\phi}^{(\nu=0)}$) is set as 
\begin{align}
  \bar{\phi}^{(0)} (\tau_I, r_J) = 
  \bar{\phi}_{O(D)} (\sqrt{r_I^2+\tau_J^2}).
\end{align}
Notice that, in order to properly place the zero-temperature solution $\bm{\varphi}_{O(D)}$ on the $\tau$ vs.\ $r$ plane, $r_{\text{max}}$ should be adjusted. We then compute the Fourier coefficients $\varphi_{n,I}^{(0)}$ using numerical integration with Simpson's rule to determine the initial guess $\bm{\varphi}^{(0)}$. The code for this procedure is shown below.
\begin{lstlisting}
# Interpolation of varphiO4
varphiO4 = interp1d(r, varphiO4, kind='cubic')
# Definition of parameters
rmax=...
Nr=...
Ntau=...
beta0=2*rmax
r=[0]*(Nr+1)
t=[0]*(Nr+1)
rdelta=rmax/Nr
tdelta=beta0/Nr
for i  in range (Nr+1):
    r[i]=pow(10,-6)+i*rdelta
    t[i]=-beta0/2+i*tdelta
# Putting varphiO4 on tau-r plane
barphi = np.zeros((Nr+1,Nr+1))
for i in range (Nr+1):
    for j in range (Nr+1):
        barphi[i,j]=varphiO4(pow((pow(r[i],2)+pow(t[j],2)),1/2))
# Calculating the Fourier components
Cos=np.zeros((Ntau+1,Nr+1))
varphiZero=np.zeros((Ntau+1,Nr+1))
for k in range (0,Ntau+1):
    for j in range (0,Nr+1):
        Cos[k,j]=np.cos(2*np.pi*k/(beta0)*t[j])
for k in range (0,Ntau+1):
    for i in range (0,Nr+1):
        if k == 0:
            varphiZero[k, i] = 1 / beta0 * simpson(barphi[i, :] * Cos[k, :], x=t)
        else:
            varphiZero[k, i] = 2 / beta0 * simpson(barphi[i, :] * Cos[k, :], x=t)
\end{lstlisting}

\section{Bounce in thermal bath}

Using $\bm{\varphi}^{(0)}$ as the initial guess, we calculate the Fourier coefficients of the bounce solution $\bm{\varphi}$ for $\beta = \beta_0 - \delta \beta$, $\beta_0 - 2 \delta \beta$, $\cdots$, $\beta_f$, with $\beta_f^{-1}$ being the highest temperature of interest. The step size $\delta \beta$ should be small enough for convergence. Note that, for the thin-wall case, the $\tau$-dependent solution vanishes around $\beta \sim 2 \bar{r}$, so the iterative process does not converge at this temperature.  The code for this procedure is shown below.
\begin{lstlisting}
# Definition of parameters
beta=beta0
betaf=...
varphi=varphiZero
deltabeta=...
UI=round((beta0-betaf)/deltabeta)
# Definition of functions used in Jacobian M and the vector field F
def dVF0(i, K, varphi):
def dVF(i, p, K, varphi):
def ddV(m,i,K,F):
def ddVmn(m,n,i,K,F):
def fk(x,y,z,k,i,r): 
def f0(x,y,z,i,r): 
# Itetarion
for U in range (UI):    
    print("finding the thermal bounce at beta=",round(beta))
    for I in range (0,15):
        F=[0]*((Ntau+1)*(Nr+1))
        varphiList=[0]*((Ntau+1)*(Nr+1))
        for k in range (0,Ntau+1):
                for i in range (0,Nr+1):
                        varphiList[round(k*(Nr+1))+i]=varphi[k,i]
        M=np.zeros(((Ntau+1)*(Nr+1),(Ntau+1)*(Nr+1)))
        for k in range (0,1):
                F[round(k*(Nr+1))]=3*varphi[k,0]-4*varphi[k,1]+varphi[k,2]
                F[k+round(Nr*(k+1))]=varphi[k,Nr]
                M[round(k*(Nr+1)),round(k*(Nr+1))]=3
                M[round(k*(Nr+1)),round(k*(Nr+1))+1]=-4
                M[round(k*(Nr+1)),round(k*(Nr+1))+2]=1
                M[k+round(Nr*(k+1)),k+round(Nr*(k+1))]=1
                for i in range (1,Nr):
                        for j in range (0,Ntau+1):
                                M[round(k*(Nr+1))+i,round(j*(Nr+1))+i]=-pow(rdelta,2)*ddVmn(0,j,i,Ntau,varphi)
                        F[round(k*(Nr+1))+i]=f0(varphi[k,i+1],varphi[k,i],varphi[k,i-1],i,r[i])
                        M[round(k*(Nr+1))+i,round(k*(Nr+1))+i+1]=1+(D-2)*rdelta/2/r[i]
                        M[round(k*(Nr+1))+i,round(k*(Nr+1))+i]=-2-pow(rdelta,2)*ddVmn(0,0,i,Ntau,varphi)
                        M[round(k*(Nr+1))+i,round(k*(Nr+1))+i-1]=1-(D-2)*rdelta/2/r[i]
        for k in range (1,Ntau+1):
                F[round(k*(Nr+1))]=3*varphi[k,0]-4*varphi[k,1]+varphi[k,2]
                F[k+round(Nr*(k+1))]=varphi[k,Nr]
                M[round(k*(Nr+1)),round(k*(Nr+1))]=3
                M[round(k*(Nr+1)),round(k*(Nr+1))+1]=-4
                M[round(k*(Nr+1)),round(k*(Nr+1))+1]=1
                M[k+round(Nr*(k+1)),k+round(Nr*(k+1))]=1
                for i in range (1,Nr):
                        for j in range (0,Ntau+1):
                                M[round(k*(Nr+1))+i,round(j*(Nr+1))+i]=-pow(rdelta,2)*ddVmn(k,j,i,Ntau,varphi)
                        F[round(k*(Nr+1))+i]=fk(varphi[k,i+1],varphi[k,i],varphi[k,i-1],k,i,r[i])
                        M[round(k*(Nr+1))+i,round(k*(Nr+1))+i+1]=1+(D-2)*rdelta/2/r[i]
                        M[round(k*(Nr+1))+i,round(k*(Nr+1))+i]=-2-pow(rdelta,2)*pow(2*np.pi*k/beta,2)-pow(rdelta,2)*ddVmn(k,k,i,Ntau,varphi)
                        M[round(k*(Nr+1))+i,round(k*(Nr+1))+i-1]=1-(D-2)*rdelta/2/r[i]
        varphidelta = np.linalg.solve(M, F)
        varphiList=varphiList-varphidelta
        print("norm of error=",np.linalg.norm(varphidelta),"norm of F=",np.linalg.norm(F),"iteration",I+1)
        for k in range (0,Ntau+1):
                for i in range (0,Nr+1):
                    varphi[k,i]=varphiList[round(k*(Nr+1))+i]
        if np.linalg.norm(varphidelta)<pow(10,-7):
#calculating the Euclidean action
            SE=0
            tdelta=beta/Nr
            Field = np.zeros((Nr+1,Nr+1))
            for i  in range (Nr+1):
                t[i]=-beta/2+i*tdelta
            for i in range (0,Nr+1):
                for j in range (0,Nr+1):
                    for k in range (0,Ntau+1):
                        Field[i,j]=Field[i,j]+varphi[k,i]*Cos[k,j]
            for i in range (0,Nr):
                for j in range (0,Nr):
                    SE=SE+tdelta*rdelta*4*np.pi*pow(r[i],D-2)*((Field[i+1,j]-Field[i-1,j])**2/rdelta**2/8+(Field[i,j+1]-Field[i,j-1])**2/tdelta**2/8+V(Field[i,j])-V(0))
            print("thermal bounce has been founded at beta=",round(beta),"with action=",SE)
            beta=beta-deltabeta
            break
\end{lstlisting}

\bibliographystyle{jhep}
\bibliography{ref}


\end{document}